%% file: root.tex
\documentclass[letterpaper, 10pt]{IEEEtran}

\usepackage{include}

\begin{document}

\title{Energy-Efficient Driving in Connected Corridors via Minimum Principle Control: Vehicle-in-the-Loop Experimental Verification in Mixed Fleets}

\author{
    {Tyler Ard,~ Longxiang Guo,~ Jihun Han,~ Yunyi Jia,~ Ardalan Vahidi,~ Dominik Karbowski}%
    \thanks{Tyler Ard ({\tt\footnotesize trard@g.clemson.edu}) and Ardalan Vahidi ({\tt\footnotesize avahidi@g.clemson.edu}) are with the Department of Mechanical Engineering, Clemson University, Clemson, SC 29634, USA.}%
    \thanks{Longxiang Guo ({\tt\footnotesize longxig@g.clemson.edu}) and Yunyi Jia ({\tt\footnotesize yunyij@g.clemson.edu}) are with the Department of Automotive Engineering, Clemson University, Greenville, SC 29607, USA.}%
    \thanks{Jihun Han ({\tt\footnotesize jihun.han@anl.gov}) and Dominik Karbowski ({\tt\footnotesize dkarbowski@anl.gov}) are with the Energy Systems Division, Argonne National Lab, Lemont, IL 60439, USA.}%
}

\maketitle

\begin{abstract}
Connected and automated vehicles (CAVs) can plan and actuate control that explicitly considers performance, system safety, and actuation constraints in a manner more efficient than their human-driven counterparts. In particular, eco-driving is enabled through connected exchange of information from signalized corridors that share their upcoming signal phase and timing (SPaT). This is accomplished in the proposed control approach, which follows first principles to plan a free-flow acceleration-optimal trajectory through green traffic light intervals by Pontryagin's Minimum Principle in a feedback manner. Urban conditions are then imposed from exogeneous traffic comprised of a mixture of human-driven vehicles (HVs) - as well as other CAVs. As such, safe disturbance compensation is achieved by implementing a model predictive controller (MPC) to anticipate and avoid collisions by issuing braking commands as necessary. The control strategy is experimentally vetted through vehicle-in-the-loop (VIL) of a prototype CAV that is embedded into a virtual traffic corridor realized through microsimulation. Up to 36\% fuel savings are measured with the proposed control approach over a human-modelled driver, and it was found connectivity in the automation approach improved fuel economy by up to 26\% over automation without. Additionally, the passive energy benefits realizable for human drivers when driving behind downstream CAVs are measured, showing up to 22\% fuel savings in a HV when driving behind a small penetration of connectivity-enabled automated vehicles.
\end{abstract}


%

\section{Introduction}
\begin{figure}
    \centering
    \includegraphics[width=1.00\columnwidth]{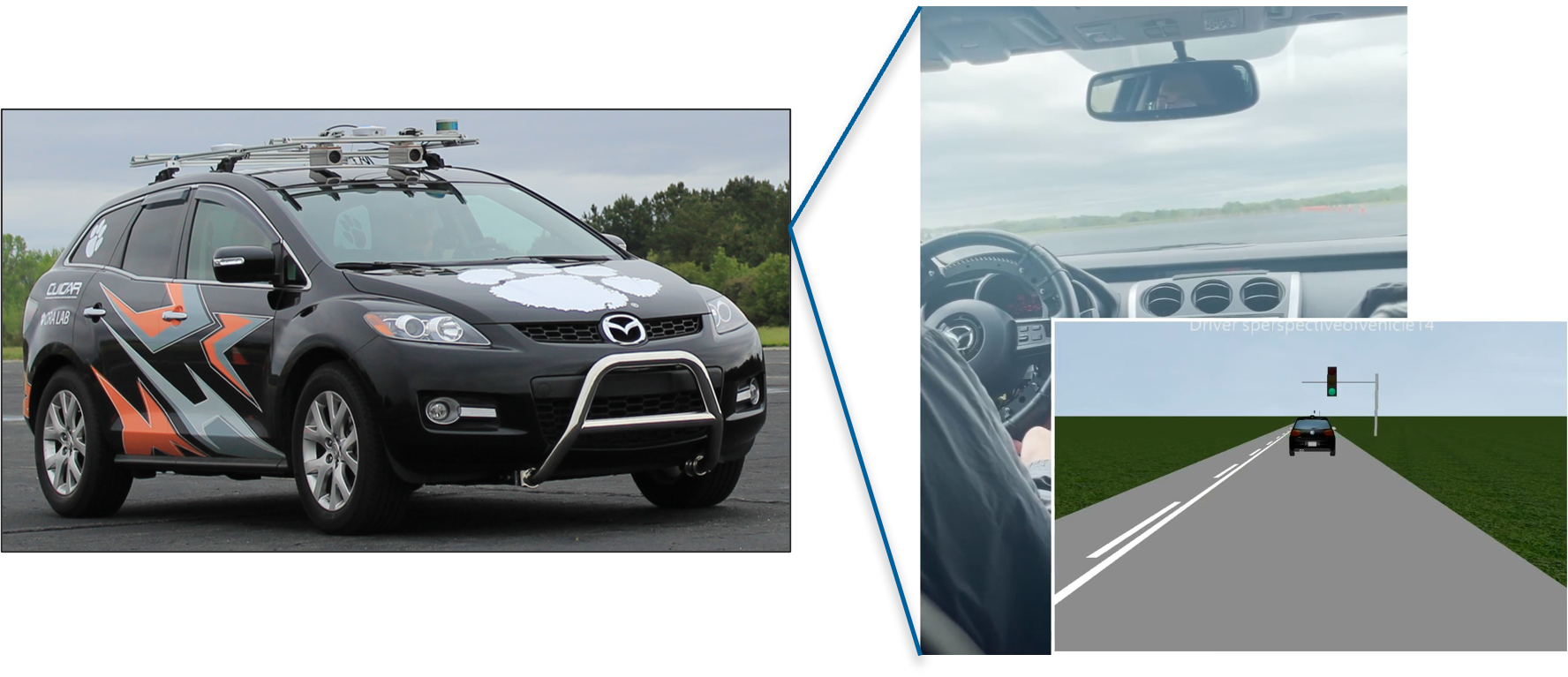}
    \caption{Viewpoint of driver in reality and in simulation during VIL testing of an urban corridor.}
    \label{fig:driversview}
\end{figure}

Cooperation in the longitudinal control of automated vehicles has soon realizable potentials for boosting energy and flow performance of real traffic in part due to connected technology \cite{Vahidi2018}. Vectors of cooperation are primarily enabled through vehicle-to-infrastructure (V2I) and vehicle-to-vehicle (V2V) communication, which both leverage a plethora of efficient mobility strategies that arise via feedback and optimal control, and have been thoroughly evaluated through representative simulation \cite{Guanetti2018}. Though, V2V-connected technologies have seen the majority share of the experimental validation to this point, even if V2I-enabled (or consequently I2V) technologies have perhaps the greater potential for environmental and mobility benefits between the two cases \cite{Wang2020}. In part, V2I technology is practically challenged by the cost of required retrofitting of existing infrastructure, though recent initiatives have looked to address the technological barriers and create validation studies \cite{Sarkar2016}.
To address this, this manuscript details the experimental performance of a prototype CAV that performs eco-driving via optimally-guided motion planning given access to the SPaT of a V2I-connected corridor of traffic lights, and contributes experimental findings in the fuel economy benefits for both automated and nearby human drivers given access to connectivity in urban settings. This is accomplished by introducing a vehicle-in-the-loop experimental architecture, in which the physical ego vehicle interacts in real-time with microsimulated vehicles and intersections in a living virtual corridor. The experimental vehicle is depicted in Figure \ref{fig:driversview}.

As connectivity and automation enters the market, significant performance benefits to an entire traffic scene are realizable under even partial connectivity and automation: while some CAVs have access to SPaT information and can directly optimize their motion, indirect performance benefits can extend to upstream traffic without said information by naturally responding to the motion of the CAV \cite{Shladover2015}. Typically, connected exchange of information allows the CAVs to anticipate downstream traffic patterns and drive in a traffic-stabilizing manner \cite{Asadi2011}. Such effects are well-studied through microsimulation of highway conditions: where increasing CAV penetrations in a merging scenario was shown to have a positive improvement on fleet fuel efficiency and traffic throughput \cite{Talebpour2016}, \cite{Rios-Torres2018}, and it was shown increasing CAV penetrations can reduce the number of braking events to suppress shockwaves that emerge in ordinary highway driving - which is directly beneficial to energy and fuel consumption \cite{ARD2020b}, \cite{Liu2020}. In urban settings, \cite{Jayawardana2022} shows a combined improvement to travel time and fuel economy in partial penetrations of CAVs equipped with learned eco-driving policies that net-benefit the entire fleet. Similarly, \cite{WAN2016} studies network-wide traffic benefits for both HVs and CAVs that are optimally guided to improve fuel economy, and \cite{Yang2017} then considers impedance due to traffic build up behind red lights. In common to many of these studies, there exists a critical penetration of CAVs before performance benefits are consistently realizable - and usually requires a larger share of the market when increasing traffic volume demands. 

Importantly, realizing benefits in connected communication does not necessarily require automation in both vehicles. Connected exchange of information from human driven vehicles in a traffic scene have also been used to boost traffic awareness in otherwise out-of-sight information, which enables more energy efficient driving strategies in CAVs. In \cite{Ge2018}, experimentation of several communication topologies of a string of connected and unconnected HVs and a CAV are employed to improve energy economy and traffic harmonization through mitigation of braking events, which is then further upscaled through microsimulation \cite{Avedisov2022}. Improved traffic harmonization and dissipation of traffic shockwaves via predictive braking are shown in \cite{Ibrahim2019} for heavy duty trucks, finding a significant fuel economy improvement. In all studies, it is suggested that the largest benefits come from scenarios with long-range communication. Additionally, V2I-enabled eco-driving algorithms can benefit human drivers by displaying suggested driving speeds during intersection approach. Passenger vehicles were shown to, on average, improve fuel economy in \cite{Mahler2017} when eco-approach speeds were suggested to the driver through a vehicle dashboard interface, whereas \cite{Chen2022} improve fuel economy of public transit buses by suggesting driving speeds to drivers.

Choice of control strategy is additionally important to realize performance benefits in CAVs. Earlier work in automated vehicle driving strategies developed predictive optimal control to anticipate surrounding vehicle and traffic light disturbances on urban roads to reduce reactionary braking - thereby improving fuel economy \cite{Kamal2011}, \cite{kamal2013}. In \cite{Wang2020b}, a combination of V2V and V2I communication is used to efficiently navigate a corridor of traffic lights, in which the strategy explicitly considers effects of ego vehicle motion on upstream traffic; and in \cite{HomChaudhuri2017} a similar concept is extended using optimal control to further boost efficiency. Analytically-based controllers are shown for energy-efficient approach to signalized and non-signalized intersections in \cite{Zhang2016}, \cite{Malikopoulos2018}. When traffic light timings are not explicitly known, \cite{Mahler2014}, \cite{Sun2020} leverage probabilistic constraints in an optimal control problem for reasoning around uncertain signal timings for more robust eco-approach to intersections.

In prototyping CAVs which rely on control strategies that leverage other connected infrastructure and vehicles, much testing burden can be placed on experimentation due to high hardware and labor costs. Vehicle-in-the-loop (VIL) testing can instead focus on instrumentation and control for a small subset of physical vehicles, and upscale the traffic scenario through virtual microsimulation to populate other agents in the traffic scenario - which would otherwise require budgeting the still-expensive automation and connectivity hardware required to produce additional CAVs and supporting infrastructure. Such VIL procedures additionally enable safe experimental testing: signal-less, autonomous intersections were first prototyped and implemented with an experimental CAV in a mixed reality fashion in \cite{Dresner2008}, \cite{Quinlan2010} - which can be otherwise dangerous to test if using multiple physical vehicles. Signalized intersections were studied in \cite{Fayazi2017} with a suggestion-based controller that advises a human driver for the best intersection approach time in a virtual urban corridor, whereas \cite{MalikopoulosVIL} prototype a motion strategy in a variety of driving maneuvers to improve fuel economy inside of a corridor consisting fully of connected and automated vehicles (that leverages VIL testing capabilities developed in \cite{FengAR}). Eco-driving strategies in a mixed-traffic highway scenario were experimented on in \cite{Ard2021} using VIL testing. 

To this point, only a limited number of experimental studies involving CAVs equipped with V2I communication have been conducted: primarily, \cite{Xia2012} novelly shows a 14\% fuel benefit due to eco-driving a CAV informed with the SPaT of a connected intersection on a traffic-less road, whereas \cite{Almannaa2019} show a 31\% fuel benefit in a single intersection when introducing surrounding traffic, and \cite{Bae2022} extend experimentation to multiple intersections in a single traffic corridor for again a 31\% fuel benefit. As such, this paper aims to further study CAV capabilities in signalized traffic corridors. Multiple routes with varied traffic intersection positions and timings are considered - including those based on real data - and notably a wider variety of scenarios are tested by additionally introducing a partial penetration of CAVs in downstream traffic to assess their energy benefits on the experimental vehicle via smoothing. Additionally, this manuscript vets the effectiveness of an analytically-derived optimal controller via application of Pontryagin's Minimum Principle - which has not seen much experimental use in vehicle motion planning, despite the technique's effectiveness and low computational and hardware demands.

\section{Upper-Level Eco-Speed Planner with Pontryagin's Minimum Principle}\label{sec:upperlevel}

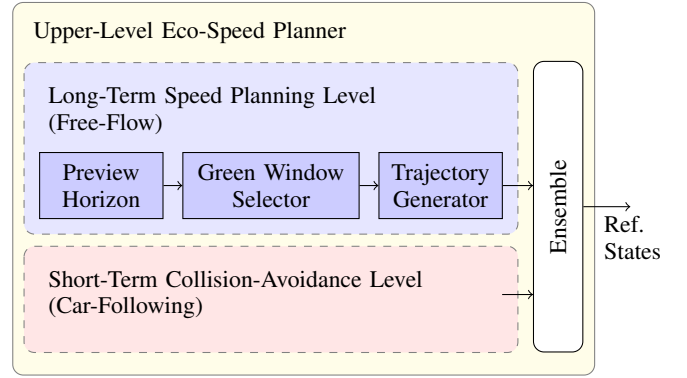
\begin{figure}
    \centering
    \input{Images/CtrlStructure}
    \caption{Upper-level eco-speed planner structure.}
    \label{fig:CTRL structure}
\end{figure}
%

The energy-efficient motion planning for the ego vehicle is performed in a two-level approach that is solved in a shrinking horizon fashion: the economic speed planning level formulates a long-term optimization problem that avoids unnecessary stops at intersections by utilizing connected traffic signal timing information, whereas the car-following level formulates a short-term optimization problem that maintains a desired time headway from the preceding vehicle. The two levels are fused through an ensemble module.

The motion planning problem is performed over a large route and contains non-convex interior constraints on the entering times through each intersection - both of which significantly increase the computational complexity. To deal with this, the optimal control problems are reformulated into bi-level optimizations, in which solutions to the newly-formed inner optimization have analytically-known explicit algebraic expressions via application of Pontryagin's Minimum Principle (PMP). The outer optimization is then efficiently, numerically solved on this condition. This approach enables control implementation even on computationally-restricted hardware, and resulted in sub-millisecond solution times for the considered scenarios.

Figure \ref{fig:CTRL structure} illustrates the upper-level controller architecture. Commanded next-step vehicle velocity and acceleration then serve as reference states to be tracked using a low-level throttle and brake controller \cite{Ard2021}. 

\subsection{Long-Term Speed Planning Level}

At the long-term speed planning level, there are three modules to: set the problem boundary conditions according to the available preview horizon, decide the feasible green light windows according to available SPaT from in-horizon intersections, and generate energy-efficient vehicle trajectories.

\subsubsection{Preview Horizon}
\begin{figure}
\centering
\includegraphics[width=1.00\columnwidth]{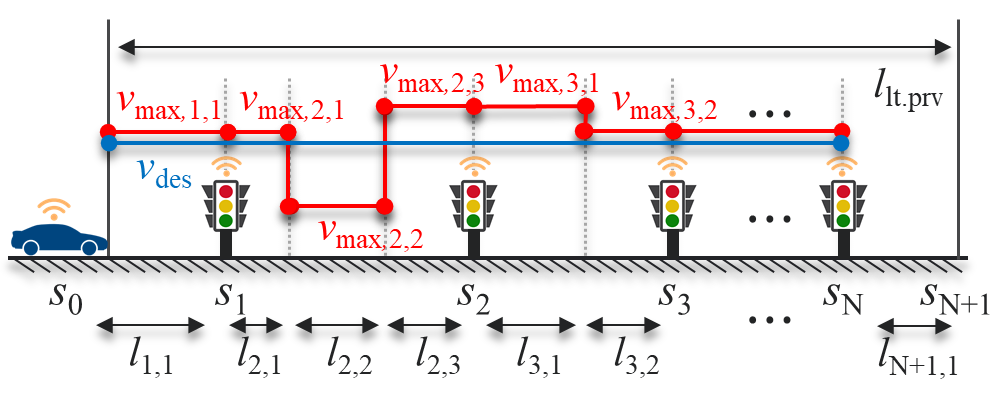}
\caption{Preview horizon diagram.}
\label{fig:PrvHrzn}
\end{figure}
The preview horizon is depicted in Figure \ref{fig:PrvHrzn}, which connects $N$ signalized intersections within range $l_\textrm{lt.prv}$. 
The end of the long-term preview horizon, $s_{N+1}$, then depends on whether a stop sign exists after the last connected traffic light. If a stop sign exists, then $s_{N+1}$ is set as its position; otherwise, a virtual point $s_{N+1}=s_{N} + \Delta s_{\textrm{vp}}$ is added.
Correspondingly, $N+1$ segments divide up the road. As the speed limit trajectory is piecewise-constant, the $i$-th road segment can consist of $n_i$ sub-segments. Accordingly, the maximum speed limit and length of the $j$-th sub-segment of the $i$-th road segment are defined as $v_{\textrm{max},i,j}$ and $l_{i,j}$, respectively.

\subsubsection{Green Window Selector}

In our previous work \cite{Han2021}, the original green window selector module selects a series of green windows for each intersection that minimize an energy objective by path-search based on Dijkstra's algorithm. We simplify this energy-optimal path search problem into a feasible path search problem so that additional computing resources are not required for any real-time implementation.

First, the \emph{selected green windows} used to enter each intersection are computed to maintain the desired travel time, while also considering the known traffic light signal phase and timing (SPaT). The minimum and desired travel times for the $i$-th road segment are defined as
\begin{equation}
    \begin{split}
    & \xi_{\textrm{min},i}= \sum^{n_i}_{j=1} \frac{l_{i,j}}{v_{\textrm{max},i,j} }, \\
    & \xi_{\textrm{des},i}= \frac{l_{i}}{v_{\textrm{des}}},  
    \end{split}
\end{equation}
where $v_{\textrm{des}}$ is the desired constant speed, and $l_{i}=s_i - s_{i-1}=\sum^{n_i}_{j=1}{l_{i, j}}$. Then, the desired entering-time at the $i$-th traffic light, $t_{\textrm{des}, i}$, can be sequentially computed starting from the current time $t_{\textrm{des},0}=t_0$. Here, note that $t_{\textrm{des},i}$ is set to the next initial green phase timing if it falls within the non-green phases. For $i=1 \ldots N{+}1$,
\begin{equation} \label{eqn:desired entering-time}
    t_{\textrm{des}.i} = 
        \begin{cases}
            t_{\textrm{des},i-1} + \xi_{\textrm{des},i} & \textrm{if } (\exists t_{\textrm{grn},i})\Phi_1(t_{\textrm{grn},i}) \\
            \{t_{\textrm{grn},i} \vert \Phi_2(t_{\textrm{grn},i}) \} + \delta t_{\textrm{grn,0}} & \textrm{else}
        \end{cases}
\end{equation}
with the two predicates
\begin{equation*}\label{eqn:predicates}
    \begin{split}
    &\Phi_1(x) = t_{\textrm{des},i-1} + \xi_{\textrm{des},i} \in [x + \delta t_{\textrm{grn,0}}, \ x + \Delta t_{\textrm{grn}} - \delta t_{\textrm{grn,f}}], \\
    &\Phi_2(x) =  t_{\textrm{des},i-1} + \xi_{\textrm{des},i} \in [x - \Delta t_{\textrm{cyc}}, \ x],
    \end{split}%
\end{equation*}
and where $t_{\textrm{grn,}i}$ is an initial green phase timing, $\Delta t_{\textrm{cyc}}$ and $\Delta t_{\textrm{grn}}$ are the durations of the entire traffic light cycle and green phase, respectively, and $\delta t_{\textrm{grn,0}}$ and $\delta t_{\textrm{grn,f}}$ are the safety time margins at the beginning and end of the green phase, respectively. 


The minimum and maximum limits, denoted \emph{feasible green windows}, are sequentially computed in a forward way (from the current time, $t_{\textrm{min},0}=t_0$) and a backward way (from the desired arrival time at the end of the preview horizon, $t_{\textrm{max},N{+}1}=t_{\textrm{des},N{+}1}$), respectively:
\begin{equation}\label{eqn:min n max limits}
    \begin{split}
        &t_{\textrm{min},i} = \max(t_{\textrm{min},i-1} + \xi_{\textrm{min},i}, \ t_{\textrm{des},i}) \ \textrm{for} \ i=1 \ldots N, \\
        &t_{\textrm{max},i} = \min(t_{\textrm{max},i+1} - \xi_{\textrm{min},i+1}, \ t_{\textrm{des},i}) \ \textrm{for} \ i=N \ldots 1.
    \end{split}
\end{equation}

In a heuristic and computationally-efficient manner, this module generates a series of feasible green windows from desired entering-times for later use in energy-efficient trajectory generation of the CAV. This is illustrated further in Figure \ref{fig:Trff_Par_study}.

\subsubsection{Reference Trajectory Generator}

The long-term-horizon optimal control problem (OCP) is formulated assuming free-flow (FF) conditions and solved to generate the energy-efficient reference trajectory. To this end, interior-point constraints are first imposed by matching vehicle positions at each intersection's entering-time, $t_{\mathrm{ent},i}$, with $s_i$.
\begin{equation}\label{Interior_point_constraint}
    s(t_{\mathrm{ent},i}) = s_i, \ i=1 \ldots N
\end{equation}
Traffic-related parameters, $\delta t_{\mathrm{des}}$ and $\delta t_{\mathrm{fea}}$, are then introduced respectively to promote entering intersections early and to tighten the feasible green windows to prevent entering intersections last-second. These parameters are calibrated to prevent long travel times within road segments that can cause unnecessary upstream traffic build-up. Compared from our previous work \cite{Han2021} which introduced a traffic-related cost with a weighting factor that must be determined from batch simulation, these parameters are simplified to be intuitive and can be calibrated manually.
\begin{equation} \label{eqn:constraint}
    \begin{split}
        {t_{\mathrm{min,n},i}} &= \max({t_{\mathrm{des},i}}-\delta t_{\mathrm{des}}, \ {t_{\mathrm{min},i}}), \\
        {t_{\mathrm{max,n},i}} &= \min({t_{\mathrm{min,n},i}}+\delta t_{\mathrm{fea}}, \ {t_{\mathrm{max},i}}). 
    \end{split}
\end{equation}
Finally, the long-term OCP is formulated by penalizing acceleration effort and using a double-integrator vehicle model.

%
\begin{equation}
    \begin{array}{rl} \label{eqn:original OCP}
    \displaystyle\min_{u} & \displaystyle J=\int_{t_0}^{t_\mathrm{f}}{\nicefrac{1}{2} \, u^2} \mathrm{d}t \\
    \textrm{s.t.} & \begin{cases}
    \ \dot{\mathbf{x}} = [x_2, \ a]^T = [x_2, \ u]^T \\
    \ x_1(t_{\mathrm{ent},i}) - s_i=0, \ i=1 \ldots N \\
    \ {t_{\mathrm{min,n},i}} \leq t_{\mathrm{ent},i} \leq t_{\mathrm{max,n},i}, \ i=1 \ldots N  \\
    \ t_{\mathrm{ent},i} - t_{\mathrm{ent},i-1} \geq \xi_{\textrm{min},i}, \ i=1 \ldots N \\
    \ \mathbf{x}(t_0) = [s_{0}, \ v_0]^T \\
    \ \mathbf{x}(t_\mathrm{f}) =  [s_\mathrm{f}, \ v_\mathrm{f}]^T
    \end{cases}
    \end{array}
\end{equation}
where $\mathbf{x}:=[s, \ v]^T$, $t_{\mathrm{ent},0}=t_0$, and boundary conditions are set by final time $t_{\mathrm{f}}=t_{\mathrm{des},N+1}$, final position $s_{\mathrm{f}}=s_{N+1}$, and final velocity $v_{\mathrm{f}}$ as either $0$ (a stop sign) or free.

The original OCP \eqref{eqn:original OCP} is then reformulated into a bi-level optimization problem to facilitate analytical treatment. The inner OCP contains the trajectory optimization of the original OCP, while the outer optimization selects intersection entering-times that minimize the running cost of the inner OCP whilst satisfying its constraints.
%
%
\begin{equation}
    \begin{array}{rl} \label{eqn:bilevel Opt}
        \displaystyle\min_{\mathbf{t}_{\mathrm{ent},i}, u^*} & \displaystyle J=\sum_{i=1}^{N+1}\int_{t_{0,i}}^{t_{\mathrm{f},i}}{\nicefrac{1}{2} \, {u^*}^2} \mathrm{d}t \\
        \textrm{s.t.} & \begin{cases}
            \ {t_{\mathrm{min,n},i}} \leq t_{\mathrm{ent},i} \leq t_{\mathrm{max,n},i}, \ i=1 \ldots N  \\
            \ t_{\mathrm{ent},i} - t_{\mathrm{ent},i-1} \geq \xi_{\textrm{min},i}, \ i=1 \ldots N \\
            \ u^*=\displaystyle \argmin_{u} \displaystyle\int_{t_0}^{t_\mathrm{f}}{\nicefrac{1}{2} \, u^2} \mathrm{d}t \\
            \quad \textrm{ s.t.} 
            \begin{cases}
                \ \dot{\mathbf{x}} = [x_2, \ u]^T \\
                \ x_1(t_{\mathrm{ent},i}) - s_i=0, \ i=1 \ldots N \\
                \ \mathbf{x}(t_0) = [s_{0}, \ v_0]^T \\
                \ \mathbf{x}(t_\mathrm{f}) =  [s_\mathrm{f}, \ v_\mathrm{f}]^T
            \end{cases}
        \end{cases} 
    \end{array}
\end{equation}
where 
\begin{equation*} \label{eqn:entering-time definition}
    t_{0,i} = 
        \begin{cases}
            t_{0} & \textrm{if } i = 1 \\
            t_{\textrm{ent},i-1} & \textrm{else},
        \end{cases} \quad
    t_{\mathrm{f},i} = 
        \begin{cases}
            t_{\mathrm{f}} & \textrm{if } i = N+1 \\
            t_{\textrm{ent},i} & \textrm{else}.
        \end{cases}
\end{equation*}

%
The conventional, numerical way to solve the inner optimization could suffer from computational burdens. To this end, the optimal control policy for the inner OCP is shown to be a piece-wise linear function over $(N+1)$ road segments through PMP analysis. The inner OCP is considered as a set of multiple unconstrained sub-OCPs with interior boundary conditions defined by $\mathrm{BC}_{\mathrm{FF},i}=(v_{i-1}, \ v_i, \ l_i, \ \xi_i)$ and with $\xi_i = t_{\mathrm{f}, i} - t_{0, i}$. These then provide the slope and intercept for each sub-OCP solution. Overall, the optimal entering speeds $\mathbf{v}^*=[v_{1}^* \ldots v_{N}^*]^T$ are given as
%
%
\begin{equation}\label{eq:equalityconstraint}
    \mathbf{v}^*=\mathbf{A}^{-1}\mathbf{y}.
\end{equation}

More details are shown in Appendices \ref{sec:MP-BVP} and \ref{sec:TP-BVP}. The cost of the outer optimization is expressed as a function of $\mathbf{v}^*$ and $\pmb{\xi}$, where $\pmb{\xi}=[\xi_{1} \ldots \xi_{N}]^T$. 
The bi-level optimization problem is finally converted into a parametric optimization problem.

%
%
\begin{equation}\label{eqn:FF optimization problem reformulation}
    \begin{array}{rl} 
        \displaystyle\min_{\pmb{\xi}} & \displaystyle J=\sum_{i=1}^{N+1}J_{i}^*\\
        \textrm{s.t.} & 
        \begin{cases}
            \ \mathbf{t_\mathrm{min,n}} \leq \mathbf{A_{\mathrm{ineq}}}\pmb{\xi} \leq \mathbf{t_\mathrm{max,n}} \\
            \ \pmb{\xi} \geq \pmb{\xi}_{\mathrm{min}} \\
            \ \mathbf{y}=\mathbf{A}\mathbf{v}^* 
        \end{cases}
    \end{array}
\end{equation}
where $J_{i}^* = {2(v_{i-1}^{*2}{+}v_{i-1}^{*}v_i^{*}{+}v_i^{*2})}{\xi_i^{-1}} - {6l_i(v_{i-1}^{*2}{+}v_i^{*2})}{\xi_i^{-2}} + {6l_i^2}{\xi_i^{-3}}$, and $\mathbf{A_{\mathrm{ineq}}}$ is a lower unitriangular matrix. 

The optimal travel times $\pmb{\xi}^*$ are obtained by numerically solving the FF parametric optimization problem \eqref{eqn:FF optimization problem reformulation} - which additionally provides $\mathbf{t}_{\mathrm{ent}}^*$. 
The equality constraint is used to compute $\mathbf{v}^*$ as in \eqref{eq:equalityconstraint}, and the optimized boundary condition of the $i$-th road segment is then defined as $\mathrm{BC}_{\mathrm{FF},i}^*=(s_{i-1}, \ v_{i-1}^*, \ {t}_{\mathrm{ent},i-1}^*, \ s_i, \ v_i^*, \ {t}_{\mathrm{ent},i}^*)$ for $i=1 \ldots N+1$. The optimal acceleration trajectory of the $i$-th road segment is computed using $\mathrm{BC}_{\mathrm{FF},i}^*$, and each segment is stitched together to finalize the entire piece-wise linear function. At every time step, the long-term speed planning level provides the free-flow acceleration of the next step $a_\mathrm{FF.n}$.

\subsection{Short-Term Car-Following Level}
The short-term car-following level focuses on car-following (CF) conditions that are not considered at the long-term speed planning level. The following CF-OCP is formulated to minimize acceleration effort while maintaining the desired inter-vehicle distance gap, $s_\mathrm{d}=v\tau_\mathrm{des} + s_\mathrm{s}$. Here, $\tau_\mathrm{des}$ is the desired time gap, and $s_\mathrm{s}$ is a standstill safety distance. 
%
%
\begin{equation}
    \begin{array}{rl} \label{eqn:original CF-OCP}
    \displaystyle\min_{u} & \displaystyle J=\int_{t_0}^{t_\mathrm{st.prv.f}}{\left[(x_1+x_2\tau_\mathrm{des} + s_\mathrm{s} - s_\mathrm{p}) + \nicefrac{1}{2} \, wu^2\right]} \mathrm{d}t \\
    &  \quad + \mu_1(x_1(t_\mathrm{st.prv.f}) - s_\mathrm{f})^2 + \mu_2(x_2(t_\mathrm{st.prv.f}) - v_\mathrm{f})^2  \\[5pt]
    \textrm{s.t.} & 
    \begin{cases}
        \ \dot{\mathbf{x}} = [x_2, \ a]^T = [x_2, \ u]^T \\
        \ x_1+x_2\tau_\mathrm{des} + s_\mathrm{s} - s_\mathrm{p} \leq 0 \\
        \ \mathbf{x}(t_0) = [s_{0}, \ v_0]^T
    \end{cases}
    \end{array}
\end{equation}
where $t_\mathrm{st.prv.f}=t_0 + t_\mathrm{st.prv}$, $t_\mathrm{st.prv}$ is a short-term preview horizon, $s_\mathrm{p}$ is the position of the preceding vehicle, and $(w, \ \mu_1, \ \mu_2)$ is a weighting factor set.

Similarly, the original CF-OCP is reformulated into a bi-level optimization problem to optimize the final conditions $s_\mathrm{f}$ and $v_\mathrm{f}$.
%
%
\begin{equation}
\begin{array}{rl} \label{eqn:bi-level CF-OPT}
    \displaystyle\min_{s_\mathrm{f}, v_\mathrm{f}, u^*} & J=\displaystyle\int_{t_0}^{t_\mathrm{st.prv.f}}{\nicefrac{1}{2} \, {u^*}^2} \mathrm{dt}  \\
    \textrm{s.t.} & \begin{cases}
        \ s_\mathrm{f}+v_\mathrm{f}\tau_\mathrm{des} + s_\mathrm{s} - s_\mathrm{p}(t_\mathrm{st.prv.f}) = 0 \\
        \ u^*= \displaystyle\argmin_{u} \displaystyle\int_{t_0}^{t_\mathrm{st.prv.f}}{\nicefrac{1}{2} \, u^2} \mathrm{d}t \\
        \quad \textrm{ s.t.} 
        \begin{cases}
            \ \dot{\mathbf{x}} = [x_2, \ u]^T \\
            \ \mathbf{x}(t_0) = [s_{0}, \ v_0]^T \\ 
            \ \mathbf{x}(t_\mathrm{st.prv.f}) =  [s_\mathrm{f}, \ v_\mathrm{f}]^T
        \end{cases}
    \end{cases} 
\end{array}
\end{equation}
%

An analytical solution to the above inner unconstrained OCP can be derived in the same manner as the inner OCP in Eq. (\ref{eqn:bilevel Opt}) and as shown in Figure \ref{fig:SummaryOfOptCtrl}. Then, $J$ is expressed as a function of $(s_\mathrm{f}, \ v_\mathrm{f})$ following $J_i^*$ as given in Eq. \eqref{eqn:FF optimization problem reformulation}, where $v_{i-1}=v_0$, $v_{i}=v_\mathrm{f}$, $l_{i}=s_\mathrm{f} - s_0$, and $\xi_{i}=t_\mathrm{st.prv}$. The bi-level CF-OCP \eqref{eqn:bi-level CF-OPT} is reformulated into a constrained parametric optimization.
%
%
\begin{equation}
    \begin{array}{lrl} \label{eqn:CF optimization problem reformulation}
        & \displaystyle\min_{s_\mathrm{f}, v_\mathrm{f}} & J=c_0+c_1s_\mathrm{f}^2+c_2s_\mathrm{f}+c_3s_\mathrm{f}v_\mathrm{f}+c_4v_\mathrm{f}+c_5v_\mathrm{f}^2\\
        & \textrm{s.t.} &  s_\mathrm{f}+\tau_\mathrm{des}v_\mathrm{f} + s_\mathrm{s} - s_\mathrm{p}(t_\mathrm{st.prv.f}) = 0
    \end{array}
\end{equation}
where $c_0=\frac{2v_0^2}{t_\mathrm{st.prv}} + \frac{6v_0s_0}{t_\mathrm{st.prv}^2} + \frac{6s_0^2}{t_\mathrm{st.prv}^3}$,  $c_1=\frac{6}{t_\mathrm{st.prv}^3}$,  $c_2=-\frac{12s_0}{t_\mathrm{st.prv}^3} - \frac{6v_0}{t_\mathrm{st.prv}^2}$,  $c_3=-\frac{6}{t_\mathrm{st.prv}^2}$,  $c_4=\frac{6s_0}{t_\mathrm{st.prv}^2} + \frac{2v_0}{t_\mathrm{st.prv}}$, and  $c_5=\frac{2}{t_\mathrm{st.prv}}$. 

To predict $s_\mathrm{p}(t_\mathrm{st.prv.f})$, we assume that the preceding vehicle (PV) maintains its current acceleration $a_\mathrm{p.0}$ but limits its speed to its maximum ($v_\mathrm{des}$) or minimum (zero).
\begin{equation}\label{eqn:PV prediction}
    \begin{split}
        & v_\mathrm{m}= \begin{cases}
            \ v_\mathrm{des} & \mathrm{if} \ a_\mathrm{p.0} \geq 0\\
            \ 0 & \mathrm{otherwise}
        \end{cases} , \\
        & t_\mathrm{m}=\min\left(t_\mathrm{st.prv}, \ \max\left(0, \ \frac{v_\mathrm{m} - v_\mathrm{p.0}}{a_\mathrm{p.0}}\right)\right), \\
        & s_\mathrm{p}(t_\mathrm{st.prv.f}) = s_\mathrm{p.0} + v_\mathrm{p.0}t_\mathrm{m} + \displaystyle \nicefrac{1}{2} \, a_\mathrm{p.0}t_\mathrm{m}^2 + v_\mathrm{m}(t_\mathrm{st.prv} {\mkern-0.1mu-\mkern-0.1mu} t_\mathrm{m})
    \end{split}
\end{equation}
where $s_\mathrm{p.0}$ and $v_\mathrm{p.0}$ are the current PV position and speed, respectively.

The Karush-Kuhn-Tucker conditions are then applied to derive the analytical solutions as \cite{Bryson}
\begin{equation}\label{eqn: vfsf solution}
\begin{split}
    \displaystyle v_\mathrm{f}^* &= -\frac{(2c_1b_1-c_3)b_0 + (c_2b_1 - c_4)}{2(c_3b_1 - c_5 - c_1b_1^2)}, \\
    s_\mathrm{f}^* &= b_0 - b_1v_\mathrm{f}^*,
\end{split}
\end{equation}
where $b_0=s_\mathrm{p}(t_\mathrm{st.prv.f}) - s_\mathrm{s}$ and $b_1=\tau_\mathrm{des}$. The optimal acceleration trajectory for the inner OCP in optimization \eqref{eqn:bi-level CF-OPT} is computed using the optimized CF boundary conditions, $\mathrm{BC}_{\mathrm{CF}}^*=(s_0, \ v_0, \ t_0, \ s_\mathrm{f}^*, \ v_\mathrm{f}^*, \ t_\mathrm{st.prv.f})$. At every time step, the short-term car-following level provides the collision-avoidance acceleration of the next step $a_\mathrm{CF.n}$.

Finally, the ensemble module limits $a_\mathrm{FF.n}$ by $a_\mathrm{CF.n}$ to determine the final reference acceleration $a_\mathrm{n}$.
\begin{equation}\label{eqn: ensemble}
    a_\mathrm{n} = \min(a_\mathrm{FF.n}, \ a_\mathrm{CF.n})
\end{equation}
Such a method can bring the energy benefits from long-sighted eco-driving as much as possible, while utilizing short-term car-following to ensure no rear-end collisions.

\section{Tunable Parameters of Eco-Speed Planner}
There are several parameters for the planner, and these parameters are pre-defined or tuned - and some can also be adjusted in real-time. In this section, key parameters are introduced to explain their impact on trajectory generation.

\subsection{Preview Horizon Length}
The long-term preview horizon for the speed planning level is distance-based - as its length $l_\textrm{lt.prv}$ is determined by the number of connected traffic lights within V2I communication range. This distance-based preview horizon is converted to a time-based preview horizon via desired arrival time in the green window selector. On the other hand, the short-term preview horizon for the car-following level is time-based, and its length is limited to avoid collisions that might be caused by prediction mismatch from the realized PV trajectory. Here, $t_\textrm{st.prv} = \unit[3]{s}$ was chosen. Note that an advanced PV prediction algorithm or intent sharing communication with a V2V-connected PV can increase $t_\textrm{st.prv}$ such that a smoother trajectory can be generated.

As the ego vehicle travels, each of the preview horizons are to be updated. 
The long-term speed planning level uses a mix of a receding horizon and shrinking horizon approach:
if there are connected traffic lights, the position of the last one defines the end of the preview horizon, and so the preview horizon is shrinking; otherwise, the preview horizon is receding with a fixed length. Here, $\Delta s_{\textrm{vp}} = \unit[100]{m}$ was set to enable eco-departure. 
On the other hand, the short-term car-following level always deploys the receding horizon methodology. 

\subsection{Desired Constant Speed}
The final arrival time at the target destination is determined in the green window selector using parameter $v_{\textrm{des}}$. In general, the final arrival time can be shortened by increasing $v_{\textrm{des}}$, while it can also be influenced by catching green phases over multiple traffic signals regardless of $v_{\textrm{des}}$. In this paper, $v_{\textrm{des}}$ is selected as a ratio to the constant maximum speed limit, $r_{\textrm{des}} = v_{\textrm{des}}/v_{\textrm{max.c}}$, where $v_{\textrm{max.c}}= l_{\textrm{lt.prv}}/\sum^{N}_{i=1}{\xi_{\textrm{min},i}}$.

\subsection{Traffic-Related Parameters}
The reference trajectory generator provides the energy-optimal intersection entering-times, but these entering-times are additionally limited by the two calibration parameters, $\delta t_{\textrm{des}}$ and $\delta t_{\textrm{fea}}$. With downstream traffic, short desired travel times can activate the car-following level frequently - which can then reduce the activity of the long-term speed planning level and reduce its energy benefits. On the other hand, long travel times can cause upstream traffic to miss available green phases or incentivize frequent lane changes. These two parameters can be intelligently adjusted using knowledge of the traffic. 

\subsection{Discussion}
%
\begin{figure}
    \centering%
    \input{Images/plannerandtraffic}
    \caption{Impact of parameters ($r_{\textrm{des}}$, $\delta t_{\textrm{des}}$, $\delta t_{\textrm{fea}}$) with tunings for high speed cases of upstream, upstream and downstream, and no traffic; and for a low speed case with no traffic.}
    \label{fig:Trff_Par_study}
\end{figure}
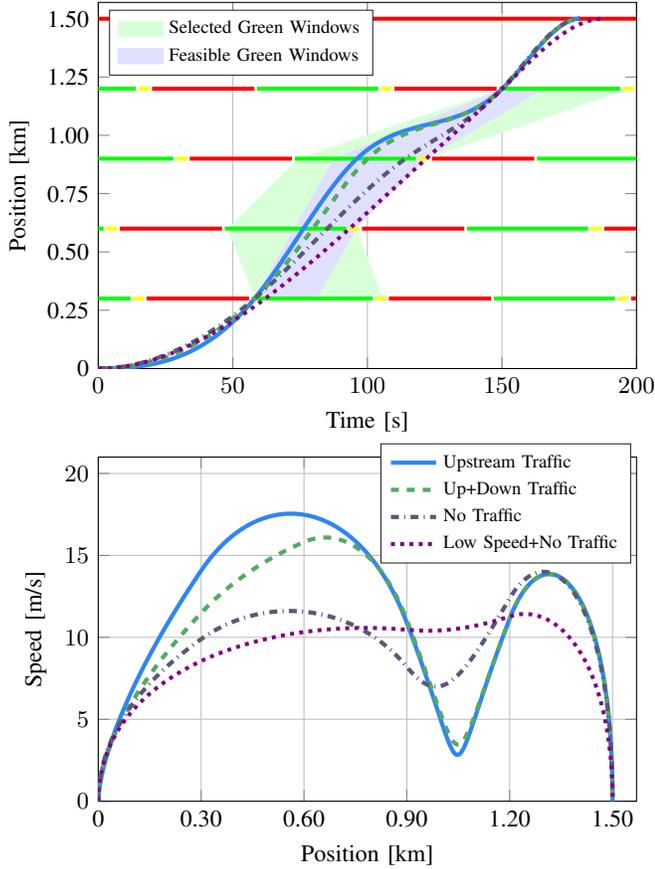

Figure \ref{fig:Trff_Par_study} shows four different trajectories depending on the values ($r_{\textrm{des}}$, $\delta t_{\textrm{des}}$, $\delta t_{\textrm{fea}}$). The first three are high-speed cases: in the case of upstream traffic, the values are set to (0.9, 5, 2) so that the ego vehicle can plan to pass through the intersections quickly. In the case of having both upstream and downstream traffic, the pair is set to (0.9, 1, 2) so that the ego vehicle can meet the original desired entering times for each intersection. In the case without traffic, the pair is set to (0.9, 5, 20) to loosen the entering-time limit constraint so that the ego vehicle can drive economically. Finally, the last case depicts a low desired speed ratio of $r_{\textrm{des}}=0.7$ under traffic-less driving which enables smoother acceleration and relaxed final travel time.




\section{Experimental Procedure}

Testing was performed by driving the ego vehicle on a closed test track with on-board computing resources that simulate nearby human vehicles (HVs), connected and automated vehicles ((C)AVs), and intersections to virtually populate a surrounding traffic scene. Specifically, a single-lane corridor is considered with 14 vehicles downstream of the ego vehicle and 4 intersections used for creating urban driving conditions. An additional red-to-green traffic light is placed at the start of the route, and a stop sign is placed at the end to create boundary conditions for each trial. In the cases where downstream traffic contains a mixture of HVs and CAVs, a CAV is placed as the leading vehicle in the string and one is placed as the vehicle 3 downstream of the ego. The total available road length is \unit[1.5]{km} at the testing facility, and a maximal vehicle velocity of \unit[17.8]{m/s} is enforced for matching urban driving conditions. 
Small performance differences (2-3\% difference in fuel economy) were measured depending on direction of travel at the testing facility due to road grade and weather conditions, so each experiment was repeated along each direction of the test track, for a total of 2 trials, and the results averaged. The vehicle engine was kept warm between each test, and ambient air temperatures were within \unit[$10^{\circ}$]{F} of each other between all tests.

\begin{figure}
    \centering
    \input{Images/scenario}
    \caption{Excerpt of scenarios considered. Diagram generated at \texttt{icograms.com}.}
    \label{fig:scenario}
\end{figure}
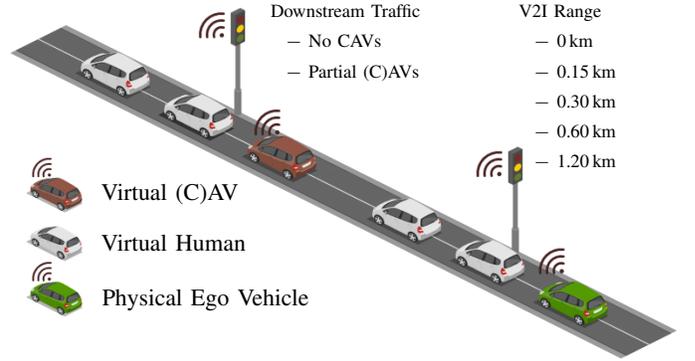

Repeatability of the experiments is enforced by creating consistent boundary conditions between each trial of an experiment. A red-to-green traffic light is assigned at the beginning of the route and a stop sign is placed at the end of the track. The first traffic light starts red so that the string of vehicles can approach and come to a full stop. Then, fuel and travel time measurements are recorded in the interval between the first traffic light turning green and the ego vehicle travelling 99\% of the distance to the end stop sign (the controllers brake at different comfort levels when approaching the stop sign). Figure \ref{fig:scenario} depicts a scaled diagram of the traffic scenario used in experiments.

The VIL experimental testing was performed for a multitude of scenarios. Namely, the downstream traffic composition was varied, the control strategy of the physical ego vehicle was varied, the maximal connectivity range of the considered V2I capability was varied, and the route that affects traffic light placement and timings was varied. The itemized list of the independent variables for experimental testing and their values are thus listed next.

\begin{itemize}
    \item The downstream traffic composition was either: 1) all humans drivers and so \emph{no CAVs}, or 2) mixed traffic with \emph{partial (C)AVs} - where positions 1 and 11 in the string are CAVs.
    \item The control strategy of the ego vehicle was: 1) an \emph{HV} using Wiedemann 99 model \cite{ptvvissim10}, 2) an unconnected \emph{AV} using car-following MPC with reactionary braking for traffic light interactions \cite{Ard2021}, or 3) a \emph{CAV} using the hierarchal eco-speed planning and car-following MPC approach of Section \ref{sec:upperlevel}.
    \item The maxmimal V2I connectivity range the intersections can broadcast at was considered for cases of \unit[0.15]{km}, \unit[0.30]{km}, \unit[0.60]{km}, and \unit[1.20]{km}. Additionally, a scenario in which there is \emph{no V2I} connectivity available was considered.
    \item The designated route affected the traffic light position and timings, and was either: 1) a \emph{synthetic} route that used a randomly chosen set of positions and SPaT of each light, or 2) the \emph{Peachtree} street route extracted from the NGSIM dataset of midtown Atlanta, Georgia, USA.
\end{itemize}

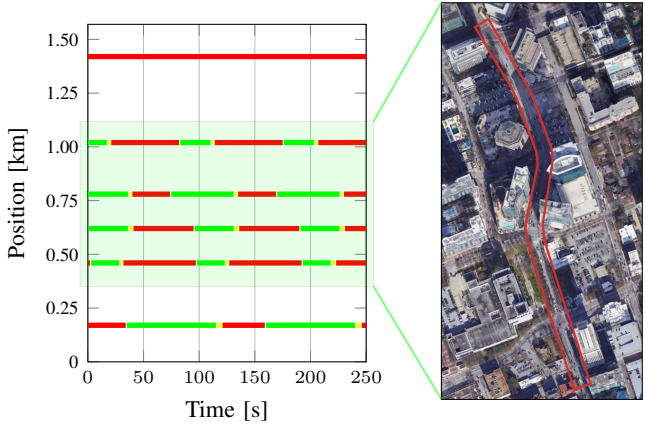
\begin{figure}
    \centering
    \input{Images/peachtree}
    \caption{Peachtree, ATL urban corridor. Signal position and green-amber-red timings shown on left, Google Earth view of corridor shown on right (outlined in red).}
    \label{fig:peachtreecorridor}
\end{figure}

Figure \ref{fig:peachtreecorridor} depicts the Peachtree urban corridor traffic light timings as green, amber, and red intervals on a position versus time diagram. Additionally, the artificial first traffic light used to start each experimental trial and the stop sign placed at the end of the track are shown.

\subsection{Vehicle Instrumentation}

\begin{figure}
    \centering
    \includegraphics[width=1.0\columnwidth]{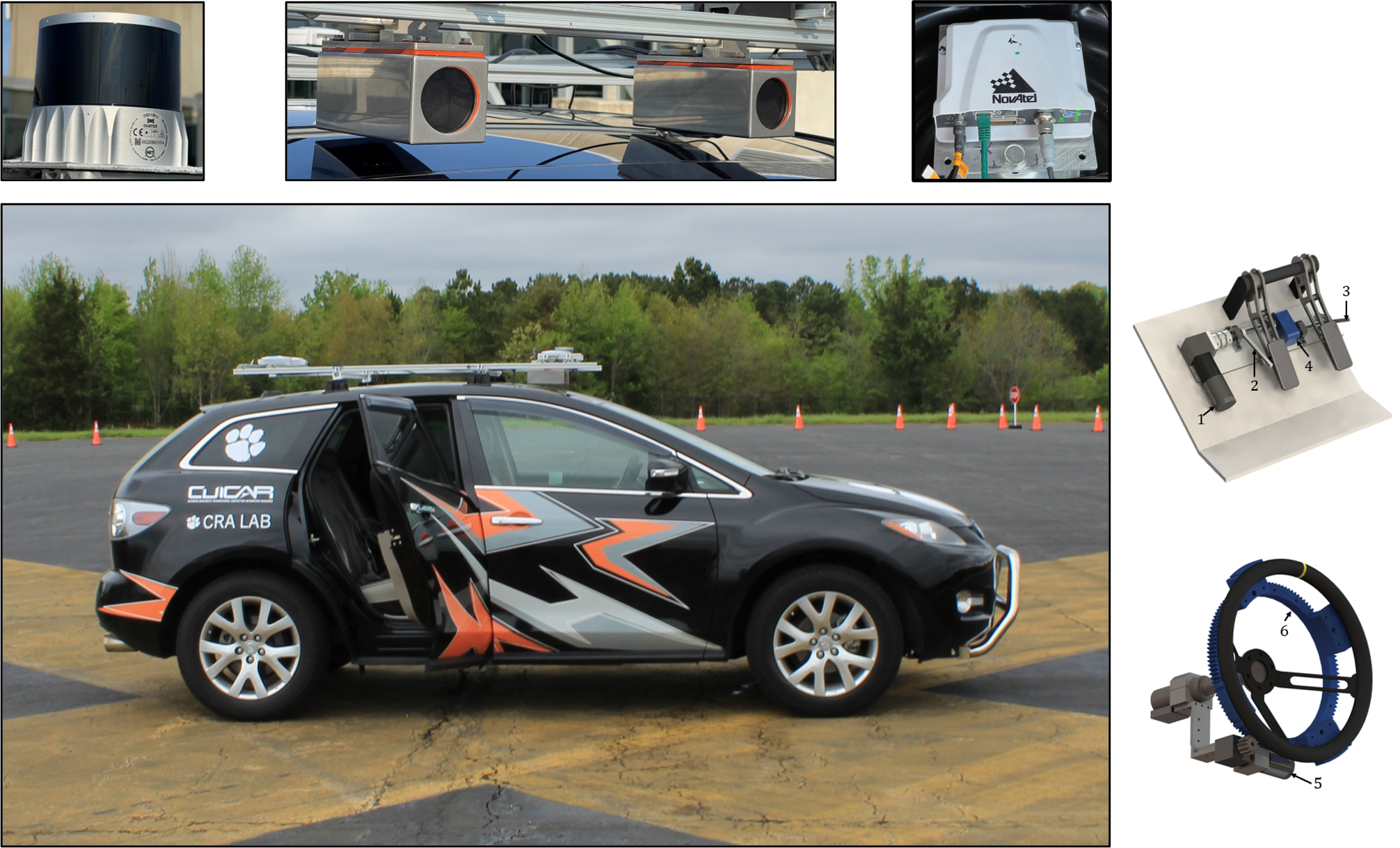}
    \caption{Prototype CAV outfit with autonomy hardware stack used in VIL testing. Field experiments performed at ITIC in Greenville, SC.}
    \label{fig:prototypecav}
\end{figure}

The experimental vehicle is a 2008 Mazda CX-7 that has been retrofit to add autonomous capabilities. The vehicle hardware stack includes a localization module, a communications and computation module, and a perception module. For localization, a Novatel PwrPak7-E1 dual-antenna RTK-GNSS with integrated IMU is used. For all onboard computation, a Dell mobile workstation with a quad-core \unit[2.8]{GHz} Intel i7 processer and \unit[16]{GB} of memory is used. The mobile workstation communicates with the sensors over ethernet through a Cisco 3560 network switch, and the clocks of the sensors are synchronized together using a TimeMachine 2000B Precision Time Protocol (PTP) server. For perception, a Velodyne VLP-16 LIDAR, an Ouster OS2-128 LIDAR, and two Mako G-319C cameras are used. It should be noted that the VIL testing procedure provides ground truth sensors for vehicle motion planning against the virtual traffic scene, and so the perception hardware is only used as an autonomous safety feature against possible real obstacles at the testing facilities - such as stray wildlife. 

Additionally, automatic vehicle actuation is achieved externally through steering wheel and pedal deflection robots using DC motors with a Roboteq MDC2460 motor driver. Experimental fuel measurements are as available in the field from OBD-II mass airflow readings that have been calibrated against fuel flow meter measurements. The vehicle used in testing is based on the experimental vehicle platform detailed previously in \cite{Ard2021}, and is depicted in Figure \ref{fig:prototypecav}.
    
\subsection{Vehicle Software}    
    

For vehicle software, socket-based communication with the microsimulation, sensor fusion, high-level motion planning, and low-level actuation control are performed through a ROS network. Position, orientation, velocity, and acceleration measurements from the guidance and inertial sensors are fused with a Kalman filter to reduce the effects of their noise on the control performance. The pedal deflection controller leverages a combined neural network-based feedforward and PID feedback control scheme to track the desired acceleration commanded from the motion planning controller of Section \ref{sec:upperlevel}. The lateral controller tracks a waypoint map - that matches the virtual microsimulated road network - using a pure pursuit scheme with look-ahead time and gain schedules for turning maneuvers of different curvatures. The low-level controller in this work executes at a frequency of \unit[35]{Hz} on the mobile workstation, whereas the high-level controller executes at a frequency of \unit[10]{Hz} (though can execute at a control frequency of over \unit[100]{Hz}).

Real-time microsimulation of the surrounding virtual traffic scene is conducted using \texttt{PTV Vissim}. The \texttt{Driver Model} interface is used to program AV and CAV model behavior for select virtual traffic in microsimulation, while the \texttt{Signal Control} interface is used to set the fixed timings of the traffic lights in simulation and to communicate their SPaT to the physical ego CAV using a server-client protocol. Primarily, VIL is enabled using the \texttt{Driving Simulator} interface to extract distance and velocity readings of surrounding vehicles, as well as distance and current state of upcoming traffic lights, execute the microsimulation, communicate with the ego vehicle, and place the ego vehicle into the microsimulation scene in real-time \cite{Ard2021}. 

\section{Results and Discussion}

\begin{figure}
    \centering
    \input{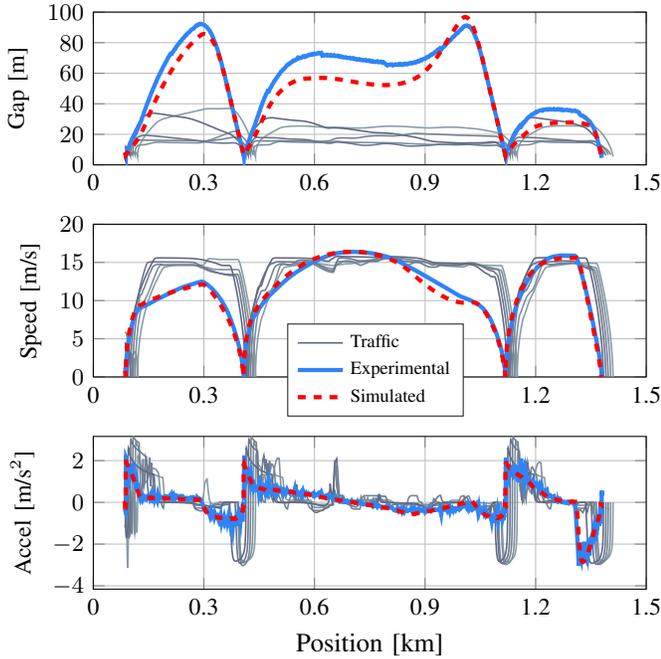}
    \caption{Experimental versus simulated trajectories of the control approach on the real vehicle. Traffic is shown in black, with darker shades representing vehicles closer to the ego.}
    \label{fig:egoconnected}
\end{figure}

For VIL experimentation, a control system on the prototype CAV is featured that processes communications from microsimulation, performs high-level motion planning for eco-driving, fuses and filters sensor information, and actuates external motors to control pedal deflection and steering angle. As such, control performance of the vehicle is first compared against a nominal simulation model of the vehicle to validate the efficacy of the deployed control strategy. Figure \ref{fig:egoconnected} depicts the realized gap, velocity, and acceleration trajectories of the ego vehicle in the same scenario from simulation and experimentation. Overall, the deployed physical vehicle control strategy strongly matches the predicted performance as from simulation.

\begin{figure}
    \centering
    \input{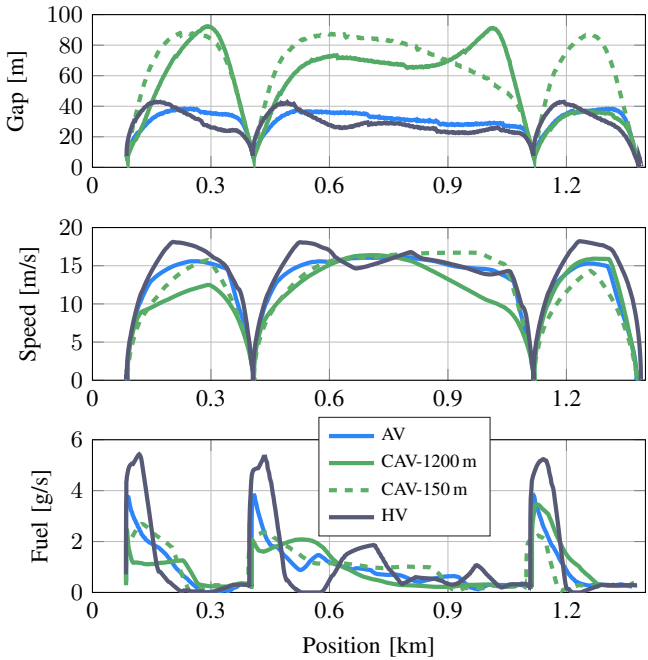}
    \caption{Performance trajectories of the control approaches.}
    \label{fig:controlsummary}
\end{figure}

Next, the different high-level controllers that perform motion planning are compared qualitatively in Figure \ref{fig:controlsummary} to examine how intersection approach and departure affected fuel performance differences. It can be seen that the CAV case with highest connectivity range cut speeding the most, which resulted in the least-incurred aerodynamic drag losses and lessened fuel expenditure. On the contrary, the HV featured the most aggressive driving patterns and reached the highest peak velocities, so fuel usage was particularly high when departing from intersections. Automation without connectivity cut unnecessary acceleration and braking for partial fuel savings and performed between the HV and CAV cases.

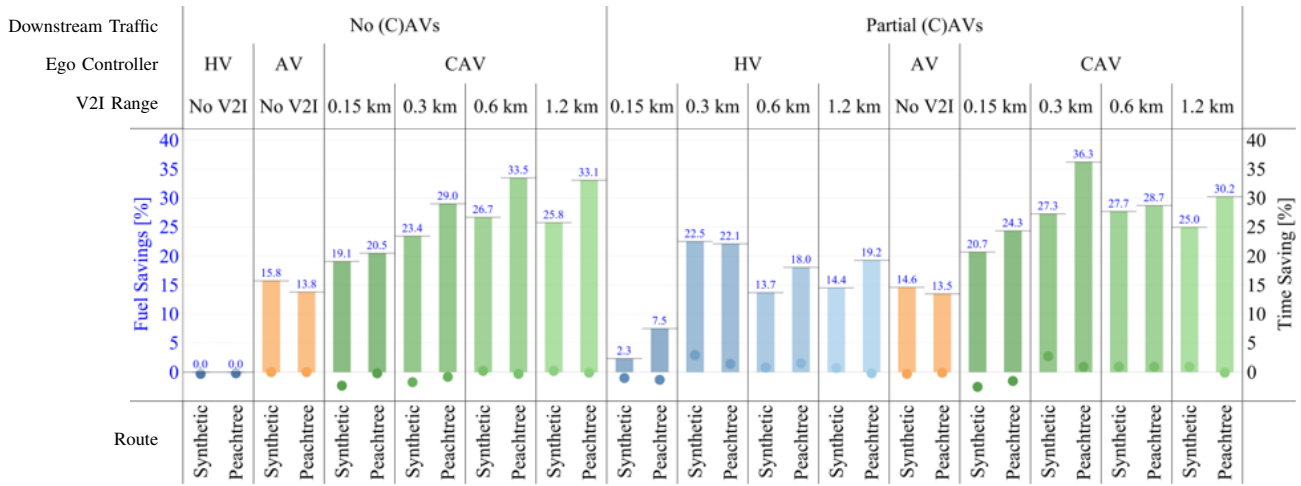
\begin{figure*}
    \centering
    \input{Images/hvbaseline}
    \caption{Measured energy (bar) and time (dot) performance of the ego vehicle control approaches compared to HV as baseline.}
    \label{fig:expresults_baseline}
\end{figure*}

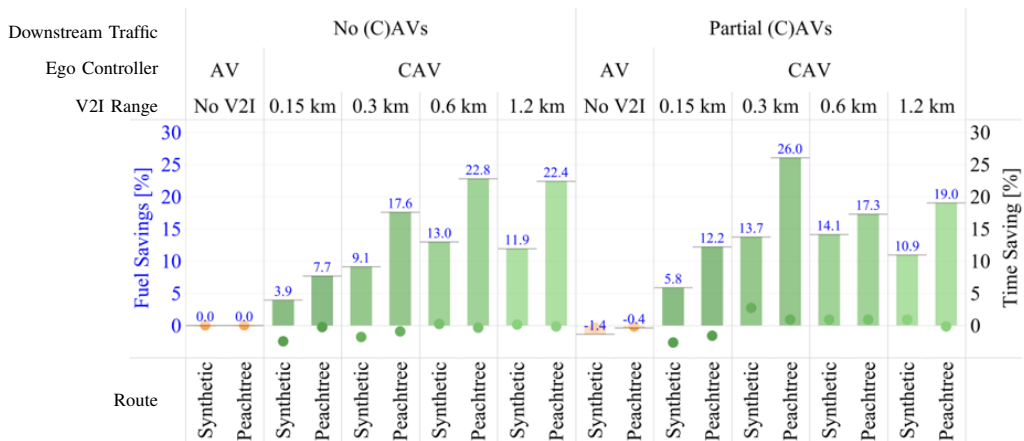
\begin{figure*}
    \centering
    \input{Images/avbaseline}
    \caption{Measured energy (bar) and time (dot) performance of the ego vehicle control approaches compared to AV as baseline.}
    \label{fig:expresults_connected}
\end{figure*}

The overall VIL experimental fuel savings and travel time improvements are summarized in Figure \ref{fig:expresults_baseline}, which reports savings for each case as compared to the HV baseline when driving in a traffic scene of all human drivers. In total, 62 experiments were run. It was found that increasing connectivity range improved fuel economy performance of both the ego vehicle when operating as a CAV and the ego vehicle when driving in mixed traffic scenarios - but was not necessarily automated or connected itself. Automation alone in the ego vehicle improved fuel economy between 13-15\% for both routes chosen - and when the ego vehicle operated with connected information as well, fuel economy was improved between 19-36\%. Travel time discrepancies were found to be within 3\% of the HV baseline case  in each trial (no more than \unit[5]{s}), and negative travel time impact due to automation was primarily found in cases where the available connectivity range was limited (\unit[0.15]{km} case). 

The presence of automation in a vehicle driving strategy has inherent energy benefits as compared to human driving behavior - due to smooth driving and reduction in unnecessary braking \cite{Ard2021}. As such, we control for automation performance and examine energy benefits found due to the V2I-connected component of a CAV driving strategy over a native AV strategy. Figure \ref{fig:expresults_connected} expresses the fuel and travel time improvements of the eco-driving approach in the CAV case with respect to the unconnected AV case. Here, the fundamental benefits of exploiting SPaT preview in eco-driving are compared to an approach that is limited to be only reactionary to traffic lights. Overall, including SPaT preview in the motion planning of a CAV improved fuel economy between 4-26\% over the preview-less AV case. When V2I connectivity range was at least \unit[300]{m} (far enough to guarantee range of at least 1 intersection when travelling throughout the route), fuel improvement increased to between 13-26\%. Travel time impact was limited to be within 3\% of the AV case in each trial (\unit[no more than 5]{s}).

For the scenarios used, there was not a significant performance difference for the ego vehicle when driving in either the partial AV case versus no AV case - largely because the downstream AV penetration rate was too lean to significantly attenuate traffic-induced speed fluctuations that require the ego vehicle to brake. However, with the same number of downstream CAVs in the network, two effects can be noted: 1) if the ego vehicle is operating as an HV, downstream CAVs provide a significant fuel benefit by indirectly improving ego traffic light interactions, and 2) if the ego vehicle is operating as a CAV with limited connectivity range, presence of downstream CAVs can further improve fuel economy - as compared to if the ego CAV has a long connectivity range, it achieves strong fuel performance regardless of the presence of downstream CAVs. 

\begin{figure}
    \centering
    \subfloat[Scenario with downstream CAVs (1200$\,$m range).\label{fig:2a}]{%
    \input{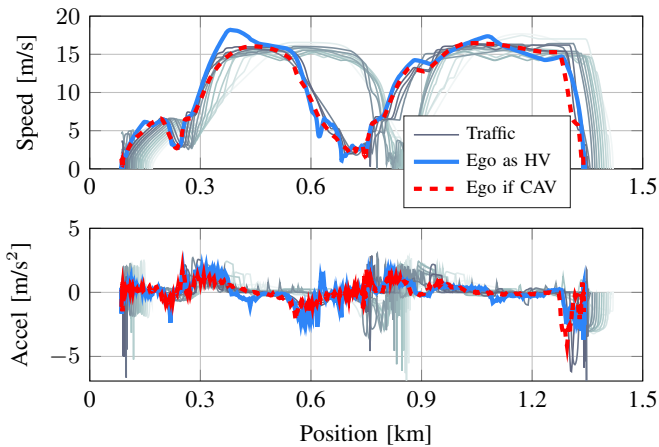}%
    }
    \\
    \subfloat[Scenario without downstream CAVs.\label{fig:2b}]{%
    \input{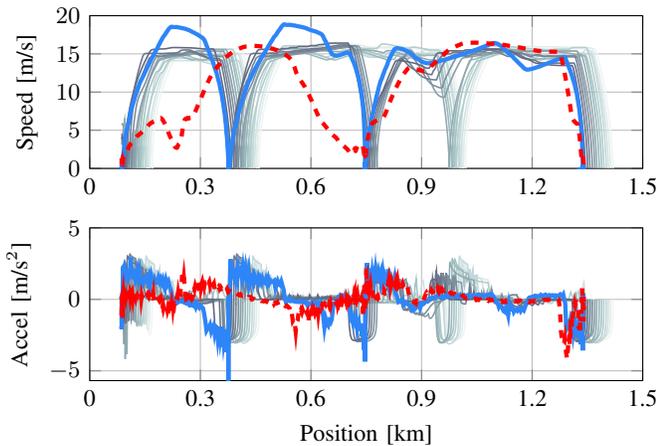}%
    }
    \caption{Performance trajectories of the ego when operated as an HV. Overlaid is the vehicle trajectory if operated as a CAV. Downstream traffic is shown in black, with darker shades representing vehicles closer to the ego.}
    \label{fig:humaneffects}
\end{figure}

\subsection{Effects on Human Drivers}

Next, it should be noted that the smooth driving of CAVs can extend significant benefits to surrounding traffic by creating a positive impedance that encourages them to drive at more eco-friendly speeds. This is particularly demonstrated in Figure \ref{fig:humaneffects}, which contrasts the velocity and acceleration profiles of the ego HV when driving in the case with partial CAVs downstream to the case with no CAVs present. Overlaid over both profiles is the realized trajectories as if the ego vehicle was driven as a CAV: it can be seen that the HV itself followed the general shape of the optimal profile by driving behind CAVs. Overall, it is important for drivers to drive near the optimal eco-speeds, however deviation can be tolerated to still realize significant fuel benefits. As can be seen in Figure \ref{fig:expresults_baseline}, the ego HV performed with up to 22\% fuel economy improvement in partial CAV scenarios. This result further depends on available connectivity range: when connectivity range is too limited, as in the \unit[0.15]{km} case, fuel savings in the ego HV are limited between 2-7\%. However, the fuel economy of the ego HV improved between 14-22\% with increasing connectivity ranges up to \unit[1.2]{km}. This study currently examines these positive benefits on surrounding traffic for single-lane scenarios, and so later studies should consider traffic impacts due to CAVs in multi-lane scenarios.

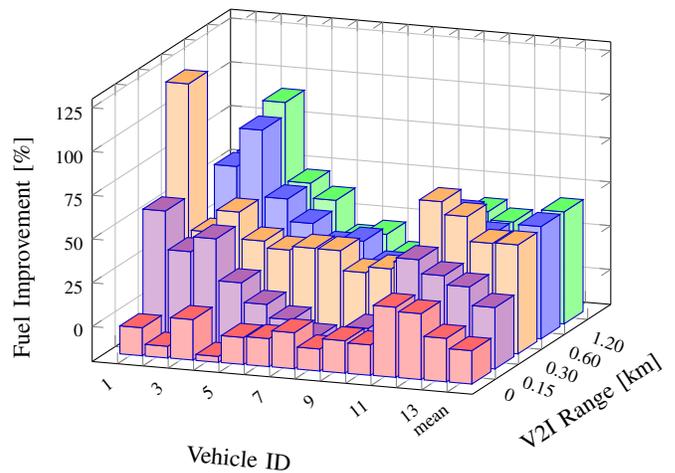
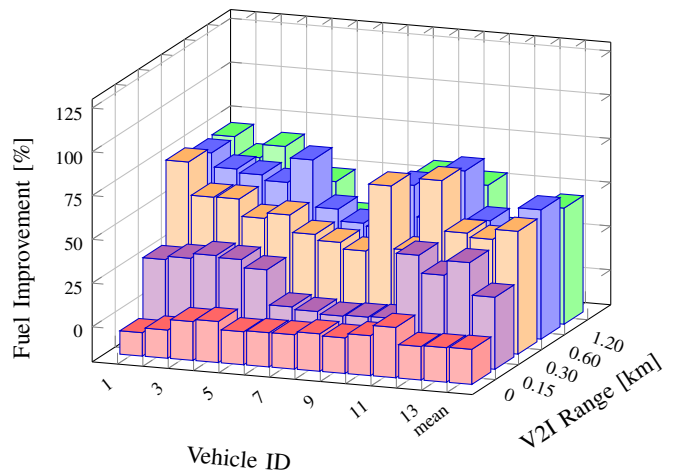
\begin{figure}
    \centering
    \subfloat[Synthetic route.\label{fig:13a}]{%
        \input{Images/autonomie}
    }
    \\
    \vspace{1em}
    \subfloat[Peachtree route.\label{fig:13b}]{%
        \input{Images/peachtreeautonomie}
    }
    \caption{Autonomie estimated fuel improvements of each simulated vehicle in the partial CAVs case with respect to its performance in the all-human scenario \cite{autonomie}. Here, vehicle IDs 1 and 11 are automated.}
    \label{fig:autonomie}
\end{figure}

Finally, the fuel performance of each microsimulated vehicle is additionally estimated. Vehicle fuel and emissions estimation software Autonomie is used to trace the microsimulated drive cycles from traffic using a mid-size SUV combustion powertrain \cite{autonomie}.
Figure \ref{fig:autonomie} depicts the estimated fuel performance of each microsimulated vehicle for the Synthetic and Peachtree routes. Here, the fuel economy improvement for each of the 13 simulated vehicles is reported with respect to its corresponding performance in the scenario with all human drivers in the fleet. Vehicle ID 1 corresponds to the leader vehicle in the string, and vehicles 1 and 11 were the designated (C)AVs during experimentation when considering the partial CAVs case. Other vehicles were designated as simulated HVs in all scenarios. It can be seen that positive fuel improvements to human drivers were concentrated to those vehicles driving more immediately behind a CAV, and in some measured cases provided significant fuel improvements of up to 80\%. It should be noted that the string fuel performance was largely unaffected in the \unit[0]{m} V2I range case (mean fuel economies were within 1\% of the baseline), and that drivers 7-9 had either neutral fuel performance in the Peachtree route or \emph{worsened} fuel performance by up to 10\% in the Synthetic route when V2I range was limited to \unit[150]{m}. Importantly, increasing V2I connectivity range had a positive fuel efficiency trend. Overall for each route, the mean fuel economy improvement of the average vehicle in the string increased between 40-50\% when connectivity range was at least \unit[300]{m}.

\subsection{Future Work}

In a system without noise or disturbances, we expect that increasing connectivity range in the CAVs will monotonically increase fuel economy. Experimentally, with traffic included in the testing scenarios, this trend was observed to be generally true except the outlier \unit[0.3]{km} cases performed the best. Traffic disrupts the planned acceleration-optimal trajectories, which become more forward-looking with higher connectivity ranges that can conversely become more sensitive to traffic impacts. The eco-speed planner of Section \ref{sec:upperlevel} assumes free-flow conditions and adjusts intersection entering times with margin parameters ($\delta t_{\mathrm{des}}$, $\delta t_{\mathrm{fea}}$). In future work, assessing traffic inhibition and queueing effects in the speed planning stage would further improve fuel economy and provide more consistently beneficial results. Additionally, the speed planner may further benefit from directly penalizing fuel consumption of the ego vehicle, and future work could investigate performance differences between the acceleration penalty and a fuel penalty. Additionally, if considering multiple lanes in an urban corridor, surrounding traffic has the opportunity to overtake the (perceivingly) slow-moving CAVs, thereby negating some of the positive energy improvements forced by driving in an energy-efficient speed profile upstream of a CAV. Multi-lane corridors are to be considered in future work.

\section{Conclusion}
The performance of eco-driving in urban corridors was presented through experimentation of a prototype connected and automated vehicle. Said eco-driving was informed by preview of upcoming SPaT as made available through V2I-connected communication with downstream traffic intersections. Eco-driving was posed as a bi-level optimization problem, in which the solutions to an inner, sub-optimization problem form the constraints on decision variables of an outer optimization. Pontryagin's Minimum Principle was utilized to derive algebraic expressions for the solution to the sub-problem: which lead to efficient, real-time compatible numerical solutions to the overall optimal control problem. Travel time was explicitly enforced as a boundary condition in the optimal control problem, and travel time differences of all trials were within \unit[5]{s} of each other.

The prototype CAV was experimentally vetted through a vehicle-in-the-loop procedure, which allowed for efficiently and quickly testing in a variety of scenarios by simulating the nearby traffic scene. The fuel and travel time performance of the ego experimental vehicle was studied when varying: downstream traffic composition between all HVs and a mixture of HVs and CAVs, available connectivity range, and route that affected placement and timing of traffic lights. Several control strategies that considered the ego vehicle as a human driver, automated driver with no SPaT preview, and a connected and automated driver with SPaT preview were considered. Overall, between 19-36\% fuel savings were measured with the proposed control approach over the human-modelled driver. 

Additionally, automation itself provides an opportunity to improve fuel economy by enforcing smooth driving and reducing unnecessary braking: between 13-15\% fuel savings were measured when using an automated vehicle strategy that was not V2I-connected. 

Downstream traffic patterns also have a role in measured fuel economy of the ego vehicle by affecting its motion. When the ego vehicle was operated as a human driver but there were V2I-connected CAVs downstream, eco-driving benefits were extended to the human driver and 14-22\% fuel savings were measured over the CAV-less scenario. Autonomie simulation was conducted to estimate the network-wide fuel usage of the simulated vehicles downstream. Energy performance of individual human drivers when driving in an all-human scenario compared to a scenario with 15\% CAVs improved by 50\% on average with a connectivity range of at least \unit[300]{m}. When driving in a scenario with AVs not equipped with connectivity, human drivers did not realize an average benefit in energy economy. Such findings suggest the potential for significant fuel savings at a network-wide level with the introduction of partial CAVs, and the critical importance that connectivity technology has on realizing said benefits. 

\appendices

\section{Multi-Point Boundary Value Problem}\label{sec:MP-BVP}
The control Hamiltonian for a dynamical system $f(x,u)$ that minimizes a performance index $J = \int_0^T L\left(x, u\right)\mathrm{d}t$ is given as $H = L + \lambda^Tf$ - and for inner OCP \eqref{eqn:bilevel Opt} is further $H = \nicefrac{1}{2} \, u^2 + \lambda_1 x_2 + \lambda_2 u$, with $\lambda_1$ and $\lambda_2$ as the position and speed co-states, respectively \cite{Bryson}.

\begin{figure}
    \centering
    \includegraphics[width=1.00\columnwidth]{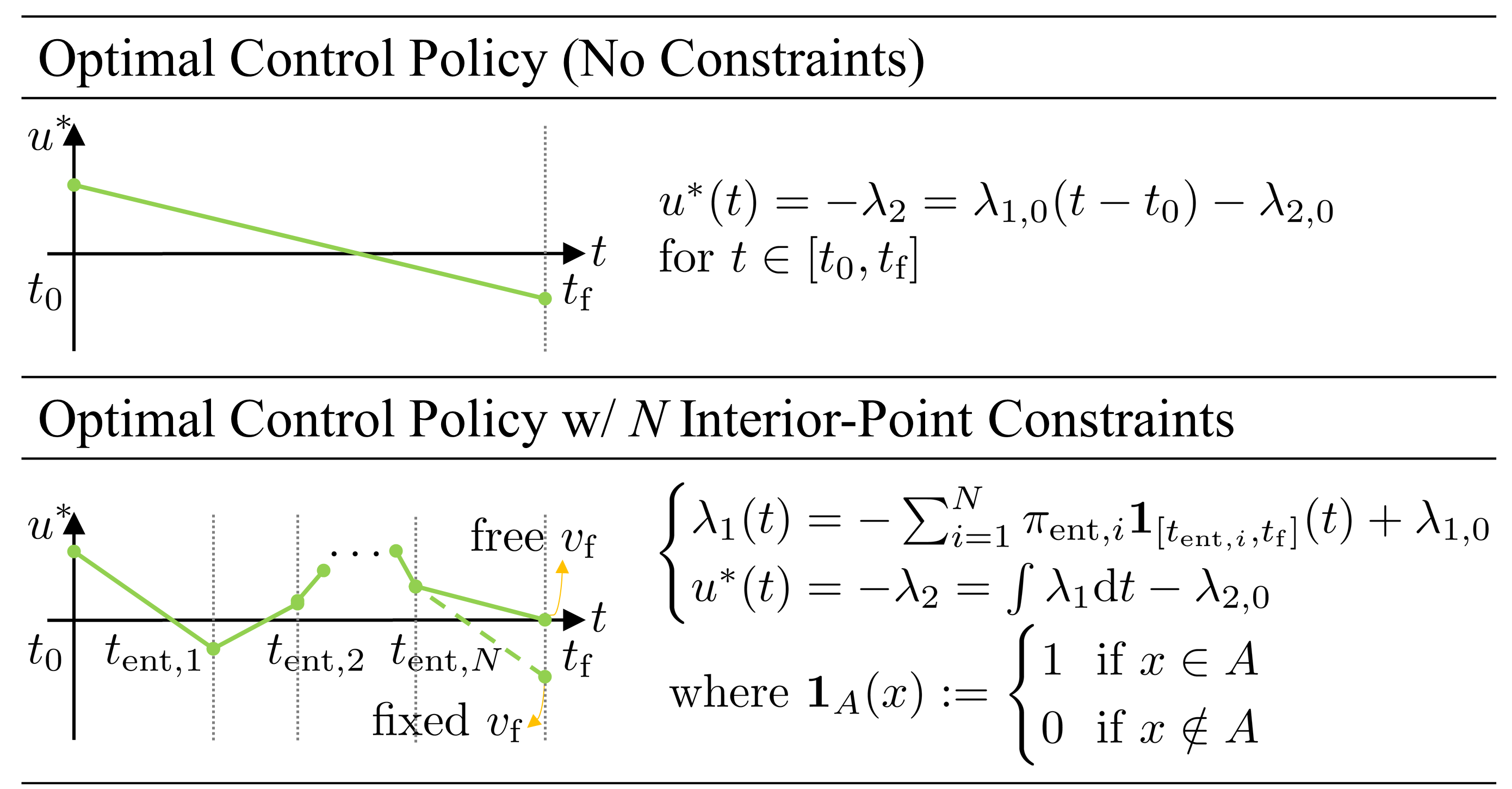}
    \caption{Summary of the optimal control policy depending on whether interior-point constraints exist.}
    \label{fig:SummaryOfOptCtrl}
\end{figure}

Deploying Pontryagin's Minimum Principle, the inner OCPs in Eq. \eqref{eqn:bilevel Opt} can be transformed into a boundary value problem (BVP), and where the interior-point constraints result in a multi-point BVP. The resulting ordinary differential equation is expressed as 
\begin{equation*}\label{eqn:ODE}
\begin{split}
    & \frac{\mathrm{d}}{\mathrm{d}t}[s^*, v^*, \lambda_1, \lambda_2]^T = [v^*, \ -\lambda_2, \ 0, \ -\lambda_1]^T \\
    & \textrm{with } \lambda_1(t_{\mathrm{ent},i}^{-}) = \lambda_1(t_{\mathrm{ent},i}^{+}) + \pi_{\mathrm{ent},i} \textrm{ for } i=1 \ldots N,
\end{split}
\end{equation*}
where $\pi_{\mathrm{ent},i}$ is a jump parameter, which leads to an optimal control policy that is piece-wise linear, as illustrated in Figure \ref{fig:SummaryOfOptCtrl}. The boundary values at $(N+2)$ points are
\begin{equation*}\label{eqn:BVP_form}
\begin{cases}
    s(t_0) = s_{0} \textrm{ and } v(t_0) = v_0, \\
    s(t_{\mathrm{ent},i}) = s_i  \textrm{ for } i=1 \ldots N, \\    
    s(t_\mathrm{f}) = s_{\mathrm{f}} \textrm{ and }  
    \begin{cases}
        v(t_\mathrm{f})=v_\mathrm{f} & \textrm{ if } v(t_\mathrm{f}) \textrm{ is fixed} \\
        \lambda_2(t_\mathrm{f})=\lambda_{2,\mathrm{f}}=0 & \textrm{ if }  v(t_\mathrm{f}) \textrm{ is free.}    
    \end{cases}
\end{cases}
\end{equation*}
We obtain a system of $(N+2)$ nonlinear equations satisfying the $(N+2)$-point boundary conditions in $(N+2)$ unknowns (such as $\pi_{\mathrm{ent},i}$ for $i=1 \ldots N+1$ and two initial values of $\lambda_1$ and $\lambda_2$).

\section{Series of Two-Point Boundary Value Problems}\label{sec:TP-BVP}

For further simplicity of computation, the multi-point BVP can then be converted to a series of two-point BVPs (TP-BVPs) in each sub-interval between points that search intersection entering-speeds and ensure the continuity of the $\lambda_2$ variable. The $i$-th TP-BVP is defined as a function of $(v_{i-1}, \ v_i, \ l_i, \ \xi_i)$, where $\xi_i=t_{\mathrm{f},i} - t_{0,i}$. 
As the optimal control policy is a linear function in time for the $i$-th TP-BVP (shown in Figure \ref{fig:SummaryOfOptCtrl}), boundary values of the $\lambda$ variables at the two points are expressed as
\begin{equation*}\label{eqn:boundary_value_lambda}
\begin{cases}
    \lambda_{1}(t_{0,i}^+)=\lambda_{1,i-1}^+=-12l_{i}\xi_{i}^{-3} + 6(v_{i-1} + v_{i})\xi_{i}^{-2}, \\
    \lambda_{2}(t_{0,i}^+)=\lambda_{2,i-1}^+ = -6l_{i}\xi_{i}^{-2} + 2(2v_{i-1} + v_{i})\xi_{i}^{-1}, \\
    \lambda_{1}(t_{\textrm{f},i}^-)=\lambda_{1,i}^-= \lambda_{1,i-1}^+, \\
    \lambda_{2}(t_{\textrm{f},i}^-)=\lambda_{2,i}^-= 6l_{i}\xi_{i}^{-2} - 2(v_{i-1} + 2v_{i})\xi_{i}^{-1}.
\end{cases}
\end{equation*}
The $i$-th continuity condition for $\lambda_2$ is imposed by the final value of
the $i$-th TP-BVP and the initial value of the $(i+1)$-th TP-BVP, as
\begin{equation*}\label{eqn:ContinuityCond}
    \lambda_{2,i}^- = \lambda_{2,i}^+ \textrm{ for } i=1 \ldots N.
\end{equation*}
In such a way, the original system of $(N+2)$ nonlinear equations can be transformed into the system of $N$ linear equations satisfying $N$ continuity conditions at $t_{\mathrm{ent},i}$ in $N$ unknowns (such as $v_i$ for $i=1 \ldots N$), as
\begin{equation*}\label{eqn:linear_system}
    \underset{\mathbf{y}\in \mathbb{R}^{N}}{%
    \vphantom{\begin{bmatrix}0\\0\\0\\0\\0\end{bmatrix}}
    \begin{bmatrix}
        y_1 - 2v_0\xi_1^{\mym}\\y_2\\ \vdots \\y_{N{\mym}}\\y_N - k_0 \\ 
        \end{bmatrix}} = 
        \underset{\pmb{A}\in \mathbb{R}^{N\times N}}{%
        \vphantom{\begin{bmatrix}0\\0\\0\\0\\0\end{bmatrix}}
        \begin{bmatrix}
        a_1 & b_1 & 0       & \cdots    & 0         \\
        b_1 & a_2 & b_2     & \cdots    & 0         \\
        0   & b_2 & a_3     & \ddots    & 0         \\
        0   & 0   & \ddots  & \ddots    & b_{N{\mym}}   \\
        0   & 0   & 0       & b_{N{\mym}}   & a_N - k_1       \\
    \end{bmatrix}}
    \underset{\mathbf{v^*}\in \mathbb{R}^{N}}{
    \begin{bmatrix}
        v_1^*\\ v_2^*\\ \vdots\\ v_{N{\mym}}^*\\ v_N^*
    \end{bmatrix}}
\end{equation*}
where $A$ is a symmetric tridiagonal matrix with $a_i= 4(\xi_i^{-1}+\xi_{i+1}^{-1})$, $b_i= 2\xi_{i+1}^{-1}$, $y_i= 6l_i\xi_i^{-2} + 6l_{i+1}\xi_{i+1}^{-2}$, and
\begin{equation*}\label{eqn:coeff_matrix}
    \begin{cases}
    (k_0, \ k_1) = (2v_\mathrm{f}\xi_1^{-1}, \ 0)             & \textrm{ if }  v(t_\mathrm{f}) \textrm{ is fixed} \\ 
    (k_0, \ k_1) = (3l_{N+1}\xi_{N+1}^{-2}, \ \xi_{N+1}^{-1}) & \textrm{ if }  v(t_\mathrm{f}) \textrm{ is free}.
    \end{cases}
\end{equation*}
\section*{Disclaimer}
The submitted manuscript has been created in part by UChicago Argonne, LLC, Operator of Argonne National Laboratory (``Argonne''). Argonne, a U.S.\ Department of Energy Office of Science laboratory, is operated under Contract No.~DE-AC02-06CH11357. The U.S.\ Government retains for itself, and others acting on its behalf, a paid-up nonexclusive, irrevocable worldwide license in said article to reproduce, prepare derivative works, distribute copies to the public, and perform publicly and display publicly, by or on behalf of the Government. The Department of Energy (DOE) will provide public access to these results of federally sponsored research in accordance with the DOE Public Access Plan.
\section*{Acknowledgment}
The authors would like to thank Mr. Rongyao (Tony) Wang, Mr. Viranjan Bhattacharrya, and Mr. Prakhar Gupta in their assistance in conducting vehicle experiments.

This report and the work described were sponsored by the U.S. Department of Energy (DOE) Vehicle Technologies Office (VTO) under the Systems and Modeling for Accelerated Research in Transportation (SMART) Mobility Laboratory Consortium, an initiative of the Energy Efficient Mobility Systems (EEMS) Program.

The following DOE Office of Energy Efficiency and Renewable Energy (EERE) managers played important roles in establishing the project concept, advancing implementation, and providing ongoing guidance: Prasad Gupte, Jacob Ward, Heather Croteau, and David Anderson.
\bibliographystyle{IEEEtran}
\bibliography{bib}

%

\begin{IEEEbiography}
[{\includegraphics[width=1in, height=1.25in, clip, keepaspectratio]{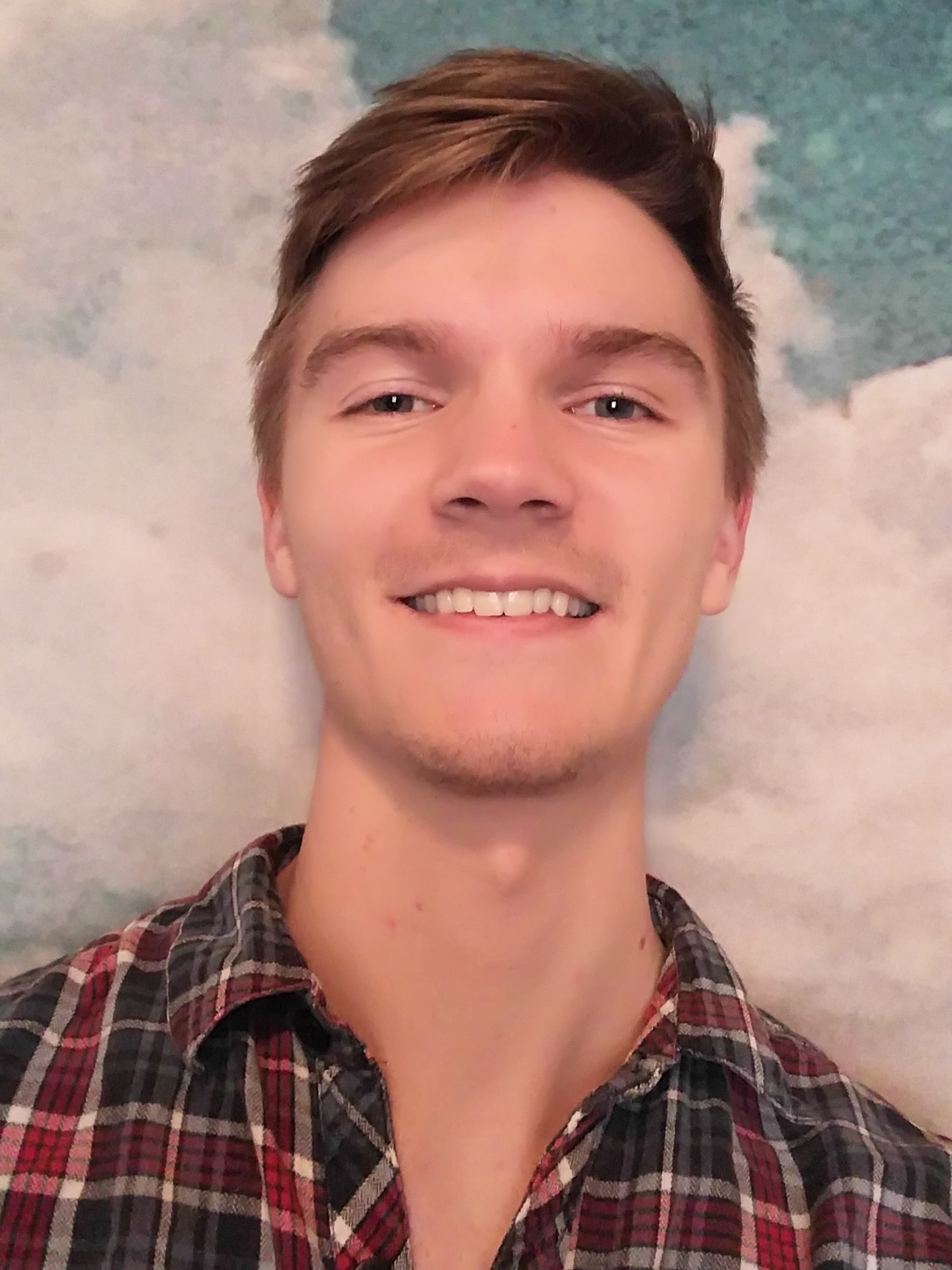}}]
{Tyler Ard}
    Tyler Ard received the B.Sc. degree in mechanical engineering from Clemson University in 2017. He is currently a candidate for the Ph.D. degree in mechanical engineering from Clemson University, where his research includes developing and verifying eco-driving strategies for ADAS and autonomous driving systems - in the aim of more efficient energy utilization of technologies already available today.
\end{IEEEbiography}

\begin{IEEEbiography}[{\includegraphics[width=1in,height=1.25in,clip,keepaspectratio]{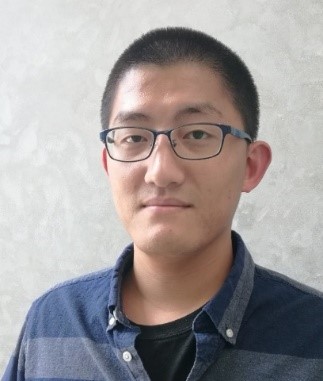}}]
{Longxiang Guo}
    Longxiang Guo (S'17-M'21) received the B.Sc. degree in mechanical engineering from Tsinghua University, Beijing, China, in 2012, the M.Sc. degree in mechanical engineering from Chinese Academy of Sciences, Shenzhen, China, in 2015, and the Ph.D. degree in automotive engineering from Clemson University in 2021. Currently, he is working as a postdoctoral fellow at Clemson University International Center for Automotive Research (CU-ICAR).
\end{IEEEbiography}

\begin{IEEEbiography}
[{\includegraphics[width=1in, height=1.25in, clip, keepaspectratio]{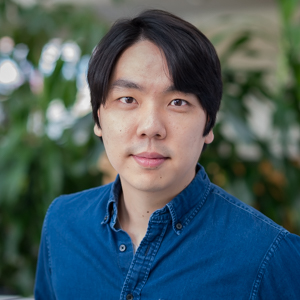}}]
{Jihun Han}
    Jihun Han received his B.Sc., M.Sc., and Ph.D. in Mechanical Engineering from Korea Advanced Institute of Science and Technology (KAIST), South Korea, in 2009, 2011 and 2016, respectively. He was a postdoctoral research associate at IFP Energies Nouvelles, France, in 2016-2017, and at Oak Ridge National Laboratory, USA, in 2017-2018. He is currently a principal research engineer at the Vehicle and Mobility Simulation group and has worked at Argonne National Laboratory since 2018. His research interests include modeling, control, and simulation with an emphasis in intelligent transportation systems, and connected and automated vehicle systems.
\end{IEEEbiography}

\vfill

\newpage

\begin{IEEEbiography}
[{\includegraphics[width=1in, height=1.25in, clip, keepaspectratio]{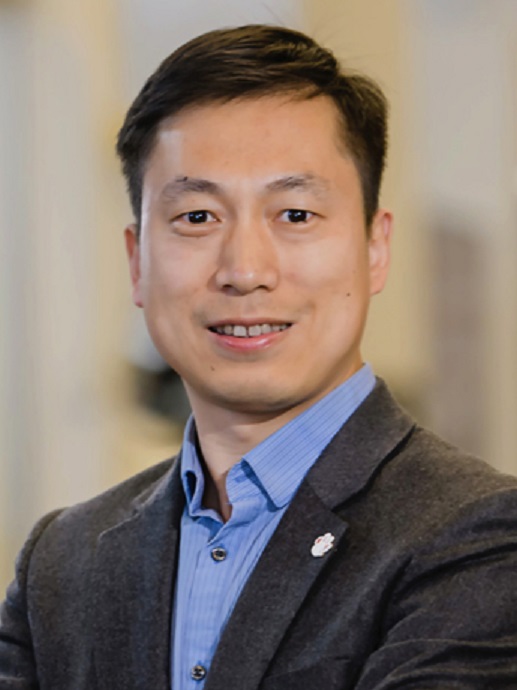}}]
{Yunyi Jia}
    Yunyi Jia (S`08-M`15-SM`20) received the B.Sc. degree from National University of Defense Technology,  M.Sc. degree from South China University of Technology and Ph.D. degree from Michigan State University. He is currently the McQueen Quattlebaum Associate Professor and the director of Collaborative Robotics and Automation (CRA) Lab in the Department of Automotive Engineering at Clemson University International Center for Automotive Research. His research interests include collaborative robotics, automated vehicles and advanced sensing systems.
\end{IEEEbiography}

\begin{IEEEbiography}
[{\includegraphics[width=1in,height=1.25in,clip,keepaspectratio]{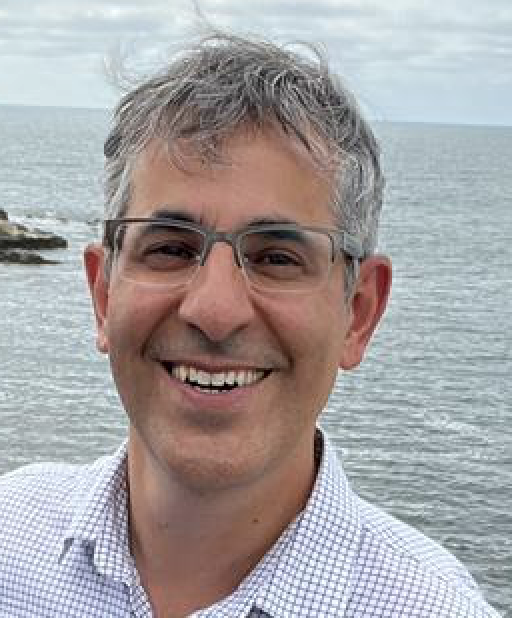}}]
{Ardalan Vahidi}
    Ardalan Vahidi (Senior Member, IEEE) received the B.Sc. and M.Sc. degrees in civil engineering from the Sharif University of Technology, Tehran, Iran, in 1996 and 1998, respectively, the M.Sc. degree in transportation safety from George Washington University, Washington, DC, USA, in 2002, and the Ph.D. degree in mechanical engineering from the University of Michigan, Ann Arbor, MI, USA, in 2005. From 2012 to 2013, he was a Visiting Scholar with the University of California at Berkeley, Berkeley, CA, USA. He has held scientific visiting positions at the BMW Technology Office, Mountain View, CA, USA, and IFP Energies Nouvelles, Rueil-Malmaison, France. He is currently Professor of Mechanical Engineering with Clemson University, Clemson, SC, USA. His recent publications span topics connected and autonomous vehicles, efficient transportation, and human performance. He is a Fellow of ASME.
\end{IEEEbiography}

\begin{IEEEbiography}[{\includegraphics[width=1in,height=1.25in,clip,keepaspectratio]{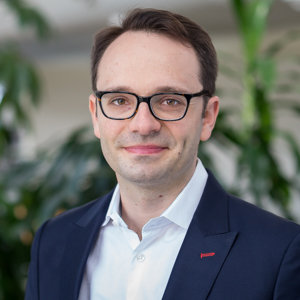}}]
{Dominik Karbowski}
    Dominik Karbowski is the Manager of the Intelligent Vehicle Controls and Low-Carbon Aviation Team, within the Vehicle and Mobility Simulation Group at Argonne National Laboratory. He leads Argonne’s effort to improve vehicle energy efficiency through control, data science, connectivity and automation, and using advanced systems simulation research. His research interests include: energy-focused controls for Connected and Automated Vehicles (CAVs), intelligent energy management and powertrain control, driver behavior, real-world driving data analytics, advanced powertrains, aircraft systems optimization. He is a major developer of simulation tools for transportation energy-efficient vehicle research. He has made significant contributions to Autonomie, Argonne’s road vehicle energy consumption tool, and has invented several software to support his research such as SVTRP, RoadRunner, and Aeronomie. He holds a master of science in engineering from Mines Paris - PSL in France.
\end{IEEEbiography}

\vfill





\end{document}

%% file: Images/CtrlStructure.tex
\pgfdeclarelayer{background}
\pgfdeclarelayer{foreground}
\pgfsetlayers{background,main,foreground}
\newcommand{\mx}[1]{\mathbf{\bm{#1}}} 
\newcommand{\vc}[1]{\mathbf{\bm{#1}}} 
%
\tikzstyle{high}=[draw, fill=blue!20, text width=4em, 
    text centered, minimum height=2.5em]
\tikzstyle{low}=[draw, fill=red!20, text width=10em, 
    text centered, minimum height=2.5em]
\tikzstyle{ensembles} = [high, text width=1.2em, fill=white!20, 
    minimum height=11em, rounded corners]
\def\blockdist{0.25}
\def\edgedist{2.5}
\begin{tikzpicture}[font=\small]%
    \node (ensemble) [ensembles] {\rotatebox{90}{Ensemble}};
    
    \path (ensemble.140)+(-0.4,0) node (trajectory) [high, anchor=east] {Trajectory Generator};
    \path (trajectory.west)+(-\blockdist,0) node (green) [high, anchor=east, text width=6em] {Green Window Selector};
    \path (green.west)+(-\blockdist,0) node (preview) [high, anchor=east] {Preview Horizon};
    
    \path[anchor=west, text width = 13em] (preview.west)+(0,1.0) node (long) {Long-Term Speed Planning Level (Free-Flow)};
    
    \path[anchor=west] (long.north west)+(-0.2,0.6) node (Upper) {Upper-Level Eco-Speed Planner};
    
    %
    
    \path[anchor=west, text width = 16em] (preview.west)+(0,-1.45) node (short) {Short-Term Collision-Avoidance Level (Car-Following)};

    \path [draw, ->] (trajectory) -- node [above] {} 
        (ensemble.west |- trajectory);
    \path [draw, ->] (green) -- node [above] {} 
        (trajectory.west |- green);
    \path [draw, ->] (preview) -- node [above] {} 
        (green.west |- preview);

    \path [draw, ->] (short)+(3.22,0) -- node [above] {} 
        (ensemble.west |- short);
        
    \path (ensemble.east)+(0.5,0) node (a) {};
    \path[draw, ->] (ensemble) -- node[text width = 2em, below right, xshift=-1.0ex] {Ref. States} (a.east |- ensemble);

    \begin{pgfonlayer}{background}
        \path (Upper.west)+(-0.15,0.4) node (a) {};
        \path (ensemble.south east)+(+0.15,-0.3) node (b) {};
        \path[fill=yellow!10,rounded corners, draw=black!50]
            (a) rectangle (b);

        \path (long.north west)+(-0.2,0.2) node (a) {};
        \path (preview.south -| trajectory.east)+(+0.2,-0.2) node (b) {};
        \path[fill=blue!10,rounded corners, draw=black!50, dashed]
            (a) rectangle (b);

        \path (short.north west)+(-0.2,0.2) node (a) {};
        \path (ensemble.south west)+(-0.2,0.0) node (b) {};
        \path[fill=red!10,rounded corners, draw=black!50, dashed]
            (a) rectangle (b);
    \end{pgfonlayer}%
\end{tikzpicture}%

%% file: Images/plannerandtraffic.tex
%
%
\definecolor{mycolor1}{rgb}{1.00000,1.00000,0.00000}%
\definecolor{mycolor2}{rgb}{0.00000,1.00000,1.00000}%
\definecolor{mycolor3}{rgb}{1.00000,0.00000,1.00000}%
\definecolor{mycolorblue}{rgb}{0.18042,0.51996,0.97333}%
\definecolor{mycolorgreen}{rgb}{0.32727,0.66364,0.40000}%
\definecolor{mycolorgreenlight}{rgb}{0.36364,0.68182,0.40000}%
\definecolor{mycolorgray}{rgb}{0.33929,0.35317,0.46429}%
\begin{tikzpicture}

\begin{axis}[%
width=0.808\columnwidth,
height=0.550\columnwidth, 
at={(0,0.650\columnwidth)},
scale only axis,
unbounded coords=jump,
xmin=0,
xmax=200,
xtick={  0,  50, 100, 150, 200, 250, 300},
xlabel style={font=\color{white!15!black}, font=\small},
xlabel={Time [s]},
ymin=-0.1,
ymax=1570,
ytick={0,  250,  500,  750, 1000, 1250, 1500, 1750, 2000, 2250, 2500, 2750, 3000},
yticklabels={0, 0.25, 0.50, 0.75, 1.00, 1.25, 1.50, 1.75, 2.00, 2.25, 2.50, 2.75, 3.00},
ylabel style={font=\color{white!15!black}, font=\small},
ylabel={Position [km]},
axis background/.style={fill=white},
xmajorgrids,
ticklabel style={font=\small},
legend style={at={(0.02, 0.98)}, anchor=north west, legend cell align=left, align=left, draw=white!15!black}
]
\addplot [color=green, line width=1.5pt, forget plot]
  table[row sep=crcr]{%
0	300\\
12	300\\
nan	nan\\
57	300\\
102	300\\
nan	nan\\
147	300\\
192	300\\
};
\addplot [color=mycolor1, line width=1.5pt, forget plot]
  table[row sep=crcr]{%
13	300\\
17	300\\
nan	nan\\
103	300\\
107	300\\
nan	nan\\
193	300\\
197	300\\
};
\addplot [color=red, line width=1.5pt, forget plot]
  table[row sep=crcr]{%
18	300\\
56	300\\
nan	nan\\
108	300\\
146	300\\
nan	nan\\
198	300\\
200	300\\
};
\addplot [color=green, line width=1.5pt, forget plot]
  table[row sep=crcr]{%
0	600\\
2	600\\
nan	nan\\
47	600\\
92	600\\
nan	nan\\
137	600\\
182	600\\
};
\addplot [color=mycolor1, line width=1.5pt, forget plot]
  table[row sep=crcr]{%
3	600\\
7	600\\
nan	nan\\
93	600\\
97	600\\
nan	nan\\
183	600\\
187	600\\
};
\addplot [color=red, line width=1.5pt, forget plot]
  table[row sep=crcr]{%
8	600\\
46	600\\
nan	nan\\
98	600\\
136	600\\
nan	nan\\
188	600\\
200	600\\
};
\addplot [color=green, line width=1.5pt, forget plot]
  table[row sep=crcr]{%
0	900\\
28	900\\
nan	nan\\
73	900\\
118	900\\
nan	nan\\
163	900\\
200	900\\
};
\addplot [color=mycolor1, line width=1.5pt, forget plot]
  table[row sep=crcr]{%
29	900\\
33	900\\
nan	nan\\
119	900\\
123	900\\
};
\addplot [color=red, line width=1.5pt, forget plot]
  table[row sep=crcr]{%
34	900\\
72	900\\
nan	nan\\
124	900\\
162	900\\
};
\addplot [color=green, line width=1.5pt, forget plot]
  table[row sep=crcr]{%
0	1200\\
14	1200\\
nan	nan\\
59	1200\\
104	1200\\
nan	nan\\
149	1200\\
194	1200\\
};
\addplot [color=mycolor1, line width=1.5pt, forget plot]
  table[row sep=crcr]{%
15	1200\\
19	1200\\
nan	nan\\
105	1200\\
109	1200\\
nan	nan\\
195	1200\\
199	1200\\
};
\addplot [color=red, line width=1.5pt, forget plot]
  table[row sep=crcr]{%
20	1200\\
58	1200\\
nan	nan\\
110	1200\\
148	1200\\
nan	nan\\
200	1200\\
};
\addplot [color=red, line width=1.5pt, forget plot]
  table[row sep=crcr]{%
0	1500\\
200	1500\\
};
\addplot [transparent, name path=A, update limits=false, forget plot]
  table[row sep=crcr]{%
58.0999999999999	300\\
48.0999999999999	600\\
74.0999999999999	900\\
150.1	1200\\
};

\addplot [transparent, name path=B, update limits=false, forget plot]
  table[row sep=crcr]{%
105.9	300\\
95.9000000000001	600\\
121.9	900\\
197.9	1200\\
};

\addplot [
    green!15!white,
    area legend,
] fill between [
    of=A and B,
    style={pattern=crosshatch}
];
\addlegendentry{\scriptsize{Selected Green Windows}}
\addplot [transparent, name path=A, update limits=false, forget plot]
  table[row sep=crcr]{%
58.0999999999999	300\\
72.8058823529411	600\\
87.5117647058823	900\\
150.1	1200\\
};

\addplot [transparent, name path=B, update limits=false, forget plot]
  table[row sep=crcr]{%
81.1941176470589	300\\
95.9000000000001	600\\
121.9	900\\
165.613903952617	1200\\
};

\addplot [
    blue!12!white,
    area legend,
] fill between [
    of=A and B,
];
\addlegendentry{\scriptsize{Feasible Green Windows}}


\addplot [color=mycolorblue, line width=1.6pt, forget plot]
  table[row sep=crcr]{%
0.0000	0.0000\\
1.0793	0.0299\\
1.9786	0.1042\\
2.8780	0.2287\\
3.7774	0.4083\\
4.6768	0.6476\\
5.5762	0.9517\\
6.4755	1.3252\\
7.1950	1.6774\\
7.9146	2.0796\\
8.6341	2.5344\\
9.3536	3.0443\\
10.0731	3.6116\\
10.7926	4.2389\\
11.5121	4.9287\\
12.2316	5.6835\\
12.9511	6.5057\\
13.6706	7.3978\\
14.3901	8.3623\\
15.1096	9.4016\\
15.8291	10.5183\\
16.7285	12.0267\\
17.6279	13.6647\\
18.5272	15.4371\\
19.4266	17.3488\\
20.3260	19.4045\\
21.2254	21.6092\\
22.1248	23.9677\\
23.0241	26.4848\\
23.9235	29.1654\\
24.8229	32.0143\\
25.7223	35.0363\\
26.6217	38.2363\\
27.5211	41.6192\\
28.4204	45.1898\\
29.3198	48.9529\\
30.2192	52.9133\\
31.2985	57.9332\\
32.3777	63.2526\\
33.4570	68.8799\\
34.5362	74.8236\\
35.6155	81.0918\\
36.6947	87.6931\\
37.7740	94.6358\\
38.8533	101.9282\\
39.9325	109.5788\\
41.0118	117.5958\\
42.0910	125.9877\\
43.1703	134.7629\\
44.2495	143.9296\\
45.5087	155.1302\\
46.7678	166.8885\\
48.0269	179.2177\\
49.2861	192.1312\\
50.5452	205.6423\\
51.8043	219.7641\\
53.0635	234.5101\\
54.3226	249.8935\\
55.5817	265.9276\\
56.8409	282.6257\\
58.1000	300.0010\\
59.5379	320.6710\\
60.9757	342.1603\\
62.5933	367.2141\\
64.2109	393.0867\\
66.0082	422.6537\\
68.1650	459.0572\\
70.8610	505.5816\\
75.7137	590.5647\\
79.1191	649.7521\\
81.4479	689.3696\\
83.4184	722.0516\\
85.2098	750.9071\\
86.8220	776.0465\\
88.2551	797.6359\\
89.6882	818.4295\\
90.9421	835.9080\\
92.1961	852.6618\\
93.4500	868.6372\\
94.5249	881.6700\\
95.5997	894.0574\\
96.6749	905.7730\\
97.7506	916.8099\\
98.8263	927.1916\\
99.7227	935.3627\\
100.6191	943.1144\\
101.5155	950.4629\\
102.4119	957.4245\\
103.3082	964.0153\\
104.2046	970.2517\\
105.1010	976.1497\\
105.9974	981.7256\\
106.8938	986.9956\\
107.7902	991.9760\\
108.6866	996.6829\\
109.5830	1001.1326\\
110.4794	1005.3412\\
111.3758	1009.3250\\
112.2722	1013.1002\\
113.3479	1017.3779\\
114.4235	1021.4066\\
115.4992	1025.2143\\
116.5749	1028.8290\\
117.8298	1032.8396\\
119.0848	1036.6701\\
120.6983	1041.4020\\
123.0289	1048.0331\\
125.5388	1055.1987\\
126.9730	1059.4379\\
128.2280	1063.2994\\
129.4829	1067.3493\\
130.5586	1071.0045\\
131.6343	1074.8590\\
132.7100	1078.9407\\
133.6064	1082.5359\\
134.5027	1086.3246\\
135.3991	1090.3229\\
136.2955	1094.5472\\
137.1919	1099.0136\\
138.0883	1103.7383\\
138.9847	1108.7376\\
139.8811	1114.0276\\
140.7775	1119.6246\\
141.6739	1125.5448\\
142.5703	1131.8044\\
143.4667	1138.4196\\
144.3631	1145.4066\\
145.2595	1152.7817\\
146.1559	1160.5610\\
147.0523	1168.7607\\
147.9487	1177.3972\\
148.8450	1186.4865\\
149.7414	1196.0449\\
150.8092	1208.0597\\
151.8730	1220.6677\\
153.1141	1236.0715\\
154.5325	1254.4299\\
156.1282	1275.8091\\
158.4331	1307.4969\\
162.1564	1358.7292\\
163.7521	1379.9580\\
165.1706	1398.1299\\
166.4117	1413.3297\\
167.4755	1425.7325\\
168.5393	1437.4702\\
169.4258	1446.6834\\
170.3123	1455.3289\\
171.0215	1461.8042\\
171.7307	1467.8607\\
172.4399	1473.4738\\
173.1491	1478.6188\\
173.6810	1482.1556\\
174.2129	1485.4048\\
174.7448	1488.3562\\
175.2767	1490.9992\\
175.8086	1493.3235\\
176.1632	1494.6909\\
176.5178	1495.9089\\
176.8724	1496.9745\\
177.2270	1497.8846\\
177.5816	1498.6361\\
177.9362	1499.2259\\
178.2908	1499.6511\\
178.4681	1499.8009\\
178.6454	1499.9084\\
178.8227	1499.9731\\
179.0000	1499.9948\\
};
\addplot [color=mycolorgreen, dashed, line width=1.6pt, forget plot]
  table[row sep=crcr]{%
0.0000	0.0000\\
0.7195	0.0358\\
1.2591	0.1100\\
1.9786	0.2727\\
2.6981	0.5088\\
3.4176	0.8192\\
4.1372	1.2047\\
4.8567	1.6659\\
5.5762	2.2038\\
6.2957	2.8189\\
7.0152	3.5122\\
7.7347	4.2843\\
8.4542	5.1361\\
9.1737	6.0683\\
9.8932	7.0817\\
10.6127	8.1769\\
11.3322	9.3549\\
12.2316	10.9449\\
13.1310	12.6667\\
14.0303	14.5219\\
14.9297	16.5121\\
15.8291	18.6387\\
16.7285	20.9031\\
17.6279	23.3069\\
18.5272	25.8516\\
19.4266	28.5387\\
20.5059	31.9533\\
21.5851	35.5776\\
22.6644	39.4143\\
23.7437	43.4659\\
24.8229	47.7350\\
25.9022	52.2243\\
26.9814	56.9362\\
28.0607	61.8734\\
29.1399	67.0385\\
30.3991	73.3558\\
31.6582	79.9909\\
32.9173	86.9479\\
34.1765	94.2308\\
35.4356	101.8438\\
36.6947	109.7910\\
37.9539	118.0765\\
39.2130	126.7044\\
40.4721	135.6787\\
41.7313	145.0037\\
43.1703	156.0954\\
44.6093	167.6565\\
46.0483	179.6931\\
47.4873	192.2114\\
48.9263	205.2175\\
50.3653	218.7175\\
51.8043	232.7175\\
53.2433	247.2238\\
54.6824	262.2423\\
56.1214	277.7793\\
57.7402	295.8856\\
59.3525	314.5836\\
60.9628	333.9102\\
62.5731	353.8649\\
64.3624	376.7466\\
66.1517	400.3464\\
67.9409	424.6338\\
69.9091	452.1080\\
71.8773	480.3374\\
74.0244	511.9469\\
76.1715	544.3548\\
78.4975	580.3034\\
81.0018	619.8997\\
89.5859	756.6256\\
91.3742	783.8425\\
92.9837	807.5607\\
94.4144	827.8963\\
95.8451	847.4159\\
97.0969	863.7413\\
98.1699	877.1165\\
99.2429	889.8730\\
100.3162	901.9701\\
101.3906	913.3877\\
102.4651	924.1365\\
103.5396	934.2469\\
104.4350	942.2065\\
105.3304	949.7615\\
106.2258	956.9294\\
107.1212	963.7278\\
108.0166	970.1743\\
108.9119	976.2864\\
109.8073	982.0817\\
110.7027	987.5778\\
111.5981	992.7924\\
112.6726	998.7028\\
113.7471	1004.2634\\
114.8215	1009.5044\\
115.8960	1014.4564\\
116.9705	1019.1496\\
118.0449	1023.6145\\
119.2985	1028.5754\\
120.5520	1033.3150\\
121.9847	1038.5226\\
123.7755	1044.8239\\
128.4315	1061.0901\\
129.8641	1066.3415\\
131.1177	1071.1324\\
132.3712	1076.1568\\
133.4457	1080.6858\\
134.5202	1085.4521\\
135.5946	1090.4860\\
136.6691	1095.8181\\
137.5645	1100.5111\\
138.4599	1105.4497\\
139.3553	1110.6515\\
140.2507	1116.1342\\
141.1461	1121.9152\\
142.0415	1128.0122\\
142.9369	1134.4427\\
143.8323	1141.2244\\
144.7276	1148.3747\\
145.6230	1155.9113\\
146.5184	1163.8517\\
147.4138	1172.2135\\
148.3092	1181.0144\\
149.2046	1190.2717\\
150.1000	1200.0033\\
151.1704	1212.2639\\
152.2407	1225.1424\\
153.4895	1240.8351\\
154.9167	1259.4856\\
156.7006	1283.5782\\
159.7333	1325.4728\\
162.2309	1359.6966\\
163.8364	1381.0236\\
165.2636	1399.2624\\
166.5123	1414.5018\\
167.5827	1426.9225\\
168.6531	1438.6614\\
169.5451	1447.8619\\
170.4370	1456.4810\\
171.1506	1462.9248\\
171.8642	1468.9399\\
172.5778	1474.5014\\
173.2914	1479.5841\\
173.8265	1483.0668\\
174.3617	1486.2554\\
174.8969	1489.1394\\
175.4321	1491.7081\\
175.9673	1493.9510\\
176.3241	1495.2600\\
176.6809	1496.4164\\
177.0377	1497.4170\\
177.3944	1498.2586\\
177.7512	1498.9381\\
178.1080	1499.4524\\
178.2864	1499.6466\\
178.4648	1499.7983\\
178.6432	1499.9072\\
178.8216	1499.9728\\
179.0000	1499.9947\\
};
\addplot [color=mycolorgray, dashdotted, line width=1.6pt, forget plot]
  table[row sep=crcr]{%
0.0000	0.0000\\
0.5396	0.0280\\
1.0793	0.1121\\
1.6189	0.2520\\
2.1585	0.4476\\
2.6981	0.6989\\
3.2378	1.0057\\
3.7774	1.3679\\
4.3170	1.7854\\
4.8567	2.2580\\
5.5762	2.9737\\
6.2957	3.7869\\
7.0152	4.6974\\
7.7347	5.7048\\
8.4542	6.8089\\
9.1737	8.0094\\
9.8932	9.3060\\
10.6127	10.6984\\
11.5121	12.5731\\
12.4115	14.5966\\
13.3108	16.7682\\
14.2102	19.0873\\
15.1096	21.5534\\
16.0090	24.1660\\
16.9084	26.9245\\
17.9876	30.4264\\
19.0669	34.1365\\
20.1461	38.0540\\
21.2254	42.1778\\
22.3046	46.5070\\
23.3839	51.0406\\
24.6430	56.5868\\
25.9022	62.4084\\
27.1613	68.5037\\
28.4204	74.8713\\
29.6796	81.5096\\
30.9387	88.4171\\
32.3777	96.6389\\
33.8167	105.2081\\
35.2557	114.1221\\
36.6947	123.3789\\
38.1337	132.9759\\
39.5728	142.9110\\
41.1916	154.4892\\
42.8105	166.4891\\
44.4294	178.9072\\
46.0483	191.7405\\
47.6672	204.9856\\
49.2861	218.6392\\
51.0848	234.2851\\
52.8836	250.4268\\
54.6824	267.0597\\
56.4811	284.1795\\
58.2787	301.7705\\
60.2449	321.5473\\
62.3897	343.7119\\
64.5346	366.4311\\
66.8582	391.5967\\
69.5393	421.2470\\
72.5778	455.4968\\
76.1526	496.4366\\
81.6935	560.6232\\
87.9585	633.0609\\
91.5446	673.8834\\
94.4135	705.9557\\
96.9237	733.4637\\
99.2547	758.4504\\
101.4064	780.9682\\
103.3788	801.0926\\
105.3511	820.6724\\
107.1442	837.9585\\
108.9372	854.7175\\
110.5510	869.3197\\
112.1647	883.4385\\
113.7785	897.0473\\
115.2106	908.6814\\
116.8209	921.3020\\
118.4312	933.5079\\
120.2204	946.6755\\
122.3674	962.0696\\
125.5880	984.7003\\
129.5242	1012.3992\\
131.6712	1027.8917\\
133.4604	1041.1761\\
135.0707	1053.5142\\
136.6810	1066.2925\\
138.1124	1078.0815\\
139.5437	1090.3310\\
140.9751	1103.0938\\
142.2275	1114.7240\\
143.4800	1126.8233\\
144.7324	1139.4273\\
145.9848	1152.5713\\
147.2373	1166.2909\\
148.4897	1180.6216\\
149.7422	1195.5988\\
150.9920	1211.2151\\
152.4191	1229.7801\\
154.0247	1251.3914\\
156.1654	1280.9857\\
161.1605	1350.3764\\
162.7660	1371.8515\\
164.1932	1390.2533\\
165.4420	1405.6809\\
166.5123	1418.3106\\
167.5827	1430.3141\\
168.4747	1439.7853\\
169.3667	1448.7278\\
170.2586	1457.0986\\
170.9722	1463.3551\\
171.6858	1469.1966\\
172.3994	1474.6012\\
173.1130	1479.5471\\
173.6481	1482.9422\\
174.1833	1486.0579\\
174.7185	1488.8847\\
175.2537	1491.4135\\
175.7889	1493.6350\\
176.1457	1494.9408\\
176.5025	1496.1032\\
176.8593	1497.1195\\
177.2160	1497.9869\\
177.5728	1498.7027\\
177.9296	1499.2642\\
178.2864	1499.6686\\
178.4648	1499.8111\\
178.6432	1499.9133\\
178.8216	1499.9748\\
179.0000	1499.9954\\
};
\addplot [color=violet, dotted, line width=1.5pt, forget plot]
  table[row sep=crcr]{%
0.0000	0.0000\\
0.5658	0.0306\\
1.1317	0.1224\\
1.6975	0.2749\\
2.2634	0.4878\\
2.8292	0.7610\\
3.3951	1.0939\\
3.9609	1.4863\\
4.5268	1.9380\\
5.0926	2.4485\\
5.6585	3.0176\\
6.4129	3.8670\\
7.1674	4.8191\\
7.9218	5.8734\\
8.6763	7.0289\\
9.4308	8.2850\\
10.1852	9.6410\\
10.9397	11.0960\\
11.6941	12.6494\\
12.6372	14.7283\\
13.5803	16.9581\\
14.5234	19.3376\\
15.4664	21.8651\\
16.4095	24.5394\\
17.3526	27.3587\\
18.4843	30.9316\\
19.6160	34.7089\\
20.7477	38.6882\\
21.8794	42.8669\\
23.0110	47.2424\\
24.1427	51.8124\\
25.4630	57.3863\\
26.7834	63.2173\\
28.1037	69.3015\\
29.4240	75.6348\\
30.7443	82.2132\\
32.2532	90.0265\\
33.7621	98.1487\\
35.2710	106.5739\\
36.7800	115.2962\\
38.2889	124.3095\\
39.9864	134.7899\\
41.6839	145.6227\\
43.3815	156.7992\\
45.0790	168.3110\\
46.9652	181.4849\\
48.8513	195.0505\\
50.7375	208.9962\\
52.8122	224.7616\\
54.8870	240.9573\\
56.9618	257.5678\\
59.2252	276.1431\\
61.4885	295.1734\\
63.9291	316.1801\\
66.3662	337.6361\\
68.9907	361.2493\\
71.6153	385.3574\\
74.4273	411.7008\\
77.4268	440.3433\\
80.4263	469.5001\\
83.6133	500.9903\\
87.1752	536.7404\\
90.9246	574.9220\\
95.2392	619.4405\\
100.1207	670.3893\\
106.5043	737.6204\\
121.9000	899.9999\\
128.6680	970.5164\\
135.6240	1043.1658\\
139.7600	1086.9220\\
143.3320	1125.2817\\
146.5280	1160.1883\\
149.5360	1193.6520\\
152.5235	1227.5131\\
158.8620	1299.5434\\
161.0991	1324.2797\\
162.9633	1344.3508\\
164.6412	1361.8772\\
166.1326	1376.9485\\
167.6240	1391.4720\\
168.9289	1403.6780\\
170.2339	1415.3682\\
171.3525	1424.9438\\
172.4710	1434.0791\\
173.5896	1442.7459\\
174.5217	1449.5898\\
175.4538	1456.0723\\
176.3859	1462.1770\\
177.3181	1467.8874\\
178.0638	1472.1609\\
178.8095	1476.1632\\
179.5552	1479.8860\\
180.3009	1483.3208\\
181.0466	1486.4593\\
181.6058	1488.6136\\
182.1651	1490.5930\\
182.7244	1492.3939\\
183.2837	1494.0128\\
183.8429	1495.4461\\
184.4022	1496.6903\\
184.7751	1497.4129\\
185.1479	1498.0489\\
185.5208	1498.5972\\
185.8936	1499.0567\\
186.2665	1499.4264\\
186.6393	1499.7052\\
187.0122	1499.8921\\
187.3850	1499.9860\\
187.5714	1499.9978\\
};
\end{axis}
\begin{axis}[%
width=0.808\columnwidth,
height=0.517\columnwidth,
at={(0.0,0.0)},
scale only axis,
xmin=-0.1,
xmax=1570,
xtick={   0,  300,  600,  900, 1200, 1500, 1800, 2100, 2400, 2700, 3000},
xticklabels={0, 0.30, 0.60, 0.90, 1.20, 1.50, 1.80, 2.10, 2.40, 2.70, 3.00, 3.30, 3.60},
xlabel style={font=\color{white!15!black}, font=\small},
xlabel={Position [km]},
ymin=0,
ymax=21,
ylabel style={font=\color{white!15!black}, font=\small},
ylabel={Speed [m/s]},
axis background/.style={fill=white},
xmajorgrids,
ymajorgrids,
ticklabel style={font=\small},
legend style={at={(0.99, 1.04)},legend cell align=left, align=left, draw=white!15!black}
]
\addplot [color=mycolorblue, line width=1.6pt]
  table[row sep=crcr]{%
0.0000	0.0000\\
0.0073	0.0273\\
0.0299	0.0566\\
0.0855	0.0986\\
0.1726	0.1441\\
0.3295	0.2058\\
0.5444	0.2728\\
0.8220	0.3453\\
1.2447	0.4394\\
1.7732	0.5412\\
2.4157	0.6507\\
3.3206	0.7884\\
4.4054	0.9365\\
5.6835	1.0953\\
7.1681	1.2645\\
8.8724	1.4444\\
10.8098	1.6348\\
12.9936	1.8357\\
15.4371	2.0472\\
18.1535	2.2692\\
21.6092	2.5359\\
25.4586	2.8164\\
29.7215	3.1107\\
34.4178	3.4187\\
39.5673	3.7405\\
45.1898	4.0761\\
51.3051	4.4255\\
57.9332	4.7887\\
65.0938	5.1657\\
72.8067	5.5564\\
81.0918	5.9609\\
89.9690	6.3792\\
99.4581	6.8113\\
109.5788	7.2572\\
120.3511	7.7169\\
131.7947	8.1903\\
143.9296	8.6775\\
156.7755	9.1785\\
170.3523	9.6933\\
184.6798	10.2219\\
199.7778	10.7643\\
215.6662	11.3204\\
232.3649	11.8904\\
249.8935	12.4741\\
268.2721	13.0716\\
287.5204	13.6828\\
305.0873	14.2267\\
312.8194	14.4523\\
320.6710	14.6699\\
328.6378	14.8797\\
339.4321	15.1469\\
350.4134	15.4001\\
361.5717	15.6392\\
372.8969	15.8641\\
384.3786	16.0750\\
396.0069	16.2718\\
407.7716	16.4544\\
419.6625	16.6229\\
431.6695	16.7774\\
443.7824	16.9177\\
455.9912	17.0439\\
468.2857	17.1560\\
480.6556	17.2540\\
493.0910	17.3379\\
505.5816	17.4077\\
518.1174	17.4634\\
530.6880	17.5050\\
543.2836	17.5325\\
555.8938	17.5459\\
568.5085	17.5451\\
581.1176	17.5303\\
593.7110	17.5013\\
606.2591	17.4584\\
618.7492	17.4014\\
631.1935	17.3305\\
643.5819	17.2455\\
655.9045	17.1465\\
668.1510	17.0335\\
680.3116	16.9065\\
692.3762	16.7655\\
704.3347	16.6105\\
716.1771	16.4414\\
727.8933	16.2583\\
739.4733	16.0613\\
750.9071	15.8502\\
762.1847	15.6251\\
773.2959	15.3859\\
784.2307	15.1328\\
794.9791	14.8657\\
802.9121	14.6561\\
810.7303	14.4387\\
818.4295	14.2133\\
826.0056	13.9801\\
833.4542	13.7390\\
840.7711	13.4901\\
847.9521	13.2332\\
854.9929	12.9685\\
861.8893	12.6958\\
868.6372	12.4153\\
875.2321	12.1270\\
881.6700	11.8307\\
887.9465	11.5265\\
894.0574	11.2145\\
899.9985	10.8946\\
911.3751	10.2594\\
923.8021	9.5512\\
935.3627	8.8785\\
947.5708	8.1531\\
960.1044	7.3924\\
972.6506	6.6158\\
990.0176	5.5239\\
1003.6856	4.6684\\
1010.8592	4.2332\\
1015.9811	3.9352\\
1020.7513	3.6727\\
1024.5938	3.4760\\
1028.2387	3.3052\\
1031.1454	3.1828\\
1033.9504	3.0785\\
1036.6701	2.9923\\
1039.3206	2.9242\\
1041.9181	2.8742\\
1044.4788	2.8422\\
1047.0191	2.8284\\
1049.0474	2.8303\\
1051.0812	2.8438\\
1053.6439	2.8770\\
1056.2445	2.9282\\
1058.8991	2.9975\\
1061.6240	3.0850\\
1064.4353	3.1905\\
1067.3493	3.3141\\
1070.3823	3.4558\\
1074.2016	3.6497\\
1078.2434	3.8696\\
1083.2776	4.1591\\
1089.5057	4.5334\\
1098.1002	5.0664\\
1135.7298	7.4196\\
1148.3092	8.1826\\
1160.5610	8.9099\\
1172.1620	9.5843\\
1184.6316	10.2942\\
1196.0449	10.9308\\
1201.9895	11.2563\\
1206.0181	11.4646\\
1210.1190	11.6643\\
1214.2893	11.8553\\
1218.5257	12.0375\\
1222.8252	12.2112\\
1227.1847	12.3761\\
1231.6012	12.5323\\
1236.0715	12.6799\\
1240.5926	12.8188\\
1245.1615	12.9490\\
1249.7749	13.0705\\
1254.4299	13.1833\\
1261.4836	13.3363\\
1268.6135	13.4697\\
1275.8091	13.5836\\
1283.0602	13.6780\\
1290.3562	13.7528\\
1297.6868	13.8081\\
1305.0417	13.8439\\
1312.4103	13.8601\\
1319.7824	13.8568\\
1327.1475	13.8339\\
1334.4953	13.7915\\
1341.8154	13.7296\\
1349.0973	13.6481\\
1356.3306	13.5471\\
1363.5051	13.4266\\
1370.6102	13.2865\\
1377.6357	13.1269\\
1382.2699	13.0097\\
1386.8610	12.8837\\
1391.4059	12.7491\\
1395.9015	12.6058\\
1400.3448	12.4538\\
1404.7326	12.2931\\
1409.0620	12.1238\\
1413.3297	11.9457\\
1417.5327	11.7590\\
1421.6680	11.5636\\
1425.7325	11.3595\\
1429.7231	11.1467\\
1433.6367	10.9253\\
1437.4702	10.6951\\
1441.2205	10.4563\\
1444.8847	10.2088\\
1448.4595	9.9526\\
1451.9420	9.6878\\
1455.3289	9.4142\\
1458.6174	9.1320\\
1461.8042	8.8411\\
1464.8864	8.5415\\
1467.8607	8.2332\\
1470.7242	7.9162\\
1473.4738	7.5906\\
1476.1063	7.2563\\
1478.6188	6.9133\\
1481.0081	6.5616\\
1483.2711	6.2012\\
1484.3543	6.0177\\
1485.4048	5.8321\\
1486.4222	5.6443\\
1487.4061	5.4544\\
1488.3562	5.2623\\
1489.2719	5.0680\\
1490.1531	4.8715\\
1490.9992	4.6729\\
1491.8099	4.4720\\
1492.5848	4.2691\\
1493.3235	4.0639\\
1494.0257	3.8566\\
1494.6909	3.6471\\
1495.3187	3.4354\\
1495.9089	3.2216\\
1496.4609	3.0056\\
1496.9745	2.7874\\
1497.4491	2.5671\\
1497.8846	2.3446\\
1498.2803	2.1199\\
1498.6361	1.8930\\
1498.9514	1.6640\\
1499.2259	1.4328\\
1499.4593	1.1994\\
1499.6511	0.9639\\
1499.8009	0.7262\\
1499.9084	0.4863\\
1499.9731	0.2442\\
1499.9948	0.0000\\
};
\addlegendentry{\scriptsize{Upstream Traffic}}

\addplot [color=mycolorgreen, dashed, line width=1.6pt]
  table[row sep=crcr]{%
0.0000	0.0000\\
0.0089	0.0498\\
0.0358	0.0998\\
0.0808	0.1501\\
0.1438	0.2006\\
0.2251	0.2514\\
0.3815	0.3281\\
0.5794	0.4055\\
0.8192	0.4834\\
1.1012	0.5619\\
1.5435	0.6675\\
2.0621	0.7742\\
2.6578	0.8820\\
3.3315	0.9908\\
4.2843	1.1283\\
5.3616	1.2675\\
6.5648	1.4083\\
7.8954	1.5509\\
9.6624	1.7241\\
11.6177	1.8997\\
13.7637	2.0777\\
16.1032	2.2581\\
19.0805	2.4715\\
22.3286	2.6883\\
25.8516	2.9083\\
29.6538	3.1315\\
33.7391	3.3581\\
38.7600	3.6209\\
44.1622	3.8881\\
49.9520	4.1595\\
56.1353	4.4351\\
62.7183	4.7150\\
70.6097	5.0350\\
79.0234	5.3604\\
87.9682	5.6911\\
97.4529	6.0273\\
107.4860	6.3688\\
118.0765	6.7157\\
130.5079	7.1075\\
143.6499	7.5060\\
157.5147	7.9111\\
172.1142	8.3228\\
187.4602	8.7412\\
203.5648	9.1663\\
220.4400	9.5980\\
238.0977	10.0363\\
258.4393	10.5262\\
279.7581	11.0241\\
302.0592	11.5297\\
314.5836	11.8020\\
329.5603	12.1119\\
344.9200	12.4135\\
360.6522	12.7069\\
376.7466	12.9919\\
393.1929	13.2686\\
409.9805	13.5371\\
427.0992	13.7972\\
444.5385	14.0491\\
464.8484	14.3267\\
485.5480	14.5936\\
506.6218	14.8496\\
528.0543	15.0947\\
549.8300	15.3291\\
571.9334	15.5526\\
594.3489	15.7652\\
608.5053	15.8881\\
617.0460	15.9488\\
628.4788	16.0132\\
639.9510	16.0587\\
651.4489	16.0853\\
662.9592	16.0931\\
674.4682	16.0819\\
685.9625	16.0519\\
697.4285	16.0029\\
708.8528	15.9350\\
717.3854	15.8718\\
725.8812	15.7978\\
734.3346	15.7133\\
742.7397	15.6181\\
751.0909	15.5123\\
759.3825	15.3959\\
767.6087	15.2688\\
775.7640	15.1311\\
783.8425	14.9828\\
791.8386	14.8238\\
799.7465	14.6543\\
807.5607	14.4741\\
815.2753	14.2832\\
822.8846	14.0818\\
830.3831	13.8697\\
837.7649	13.6469\\
845.0243	13.4136\\
852.1557	13.1696\\
859.1533	12.9150\\
866.0115	12.6498\\
872.7246	12.3739\\
879.2868	12.0874\\
885.6924	11.7903\\
891.9358	11.4825\\
898.0112	11.1641\\
907.7644	10.6250\\
922.3902	9.8014\\
935.8721	9.0281\\
951.2255	8.1321\\
970.1743	7.0096\\
990.7392	5.7925\\
999.6530	5.2793\\
1006.0444	4.9242\\
1012.0147	4.6075\\
1016.8335	4.3667\\
1021.4087	4.1542\\
1025.0573	3.9986\\
1028.5754	3.8627\\
1031.9806	3.7465\\
1035.2905	3.6498\\
1038.5226	3.5728\\
1041.0641	3.5253\\
1043.5762	3.4904\\
1046.0677	3.4681\\
1048.5477	3.4583\\
1051.0252	3.4611\\
1053.5092	3.4764\\
1056.0087	3.5043\\
1058.5327	3.5448\\
1061.0901	3.5978\\
1064.3477	3.6817\\
1067.6893	3.7853\\
1071.1324	3.9085\\
1074.6946	4.0514\\
1078.3935	4.2138\\
1082.2467	4.3959\\
1087.0988	4.6404\\
1092.2288	4.9130\\
1098.6054	5.2669\\
1106.4684	5.7184\\
1118.4098	6.4213\\
1154.3722	8.5494\\
1168.8171	9.3860\\
1182.8287	10.1830\\
1196.0526	10.9216\\
1202.0002	11.2480\\
1206.0510	11.4580\\
1210.1752	11.6592\\
1214.3697	11.8517\\
1218.6312	12.0353\\
1222.9566	12.2101\\
1227.3429	12.3761\\
1231.7868	12.5334\\
1236.2853	12.6818\\
1240.8351	12.8214\\
1245.4332	12.9523\\
1250.0764	13.0743\\
1254.7616	13.1875\\
1259.4856	13.2920\\
1266.6376	13.4321\\
1273.8592	13.5525\\
1281.1400	13.6530\\
1288.4693	13.7338\\
1295.8365	13.7948\\
1303.2311	13.8359\\
1310.6424	13.8573\\
1318.0598	13.8589\\
1325.4728	13.8407\\
1332.8708	13.8027\\
1340.2431	13.7448\\
1347.5791	13.6672\\
1354.8683	13.5698\\
1362.1001	13.4526\\
1369.2639	13.3156\\
1376.3491	13.1588\\
1381.0236	13.0433\\
1385.6553	12.9190\\
1390.2411	12.7858\\
1394.7779	12.6439\\
1399.2624	12.4932\\
1403.6916	12.3337\\
1408.0623	12.1653\\
1412.3714	11.9882\\
1416.6157	11.8023\\
1420.7921	11.6076\\
1424.8974	11.4041\\
1428.9286	11.1918\\
1432.8825	10.9706\\
1436.7559	10.7407\\
1440.5457	10.5020\\
1444.2487	10.2545\\
1447.8619	9.9982\\
1451.3820	9.7331\\
1454.8060	9.4592\\
1458.1308	9.1765\\
1461.3530	8.8850\\
1464.4697	8.5847\\
1467.4777	8.2756\\
1470.3738	7.9577\\
1473.1550	7.6310\\
1475.8180	7.2955\\
1478.3597	6.9512\\
1480.7770	6.5981\\
1483.0668	6.2362\\
1484.1628	6.0519\\
1485.2259	5.8655\\
1486.2554	5.6769\\
1487.2511	5.4860\\
1488.2126	5.2930\\
1489.1394	5.0977\\
1490.0312	4.9003\\
1490.8876	4.7006\\
1491.7081	4.4988\\
1492.4925	4.2947\\
1493.2403	4.0885\\
1493.9510	3.8801\\
1494.6244	3.6694\\
1495.2600	3.4566\\
1495.8575	3.2415\\
1496.4164	3.0243\\
1496.9363	2.8049\\
1497.4170	2.5832\\
1497.8578	2.3594\\
1498.2586	2.1333\\
1498.6188	1.9051\\
1498.9381	1.6747\\
1499.2161	1.4420\\
1499.4524	1.2072\\
1499.6466	0.9701\\
1499.7983	0.7309\\
1499.9072	0.4895\\
1499.9728	0.2458\\
1499.9947	0.0000\\
};
\addlegendentry{\scriptsize{Up+Down Traffic}}

\addplot [color=mycolorgray, dashdotted, line width=1.6pt]
  table[row sep=crcr]{%
0.0000	0.0000\\
0.0125	0.0693\\
0.0498	0.1385\\
0.1121	0.2075\\
0.1991	0.2765\\
0.3110	0.3454\\
0.4476	0.4142\\
0.6090	0.4828\\
0.8973	0.5857\\
1.2410	0.6883\\
1.6401	0.7907\\
2.0943	0.8928\\
2.6036	0.9947\\
3.3681	1.1303\\
4.2300	1.2654\\
5.1890	1.4002\\
6.2448	1.5346\\
7.3971	1.6685\\
8.9728	1.8354\\
10.6984	2.0016\\
12.5731	2.1673\\
14.5966	2.3323\\
16.7682	2.4967\\
19.5688	2.6931\\
22.5809	2.8886\\
25.8036	3.0833\\
29.2359	3.2770\\
33.5037	3.5019\\
38.0540	3.7255\\
42.8851	3.9480\\
47.9955	4.1692\\
53.3837	4.3892\\
59.8798	4.6391\\
66.7343	4.8874\\
73.9450	5.1341\\
81.5096	5.3792\\
90.4399	5.6531\\
99.8118	5.9249\\
109.6221	6.1947\\
119.8676	6.4624\\
130.5449	6.7282\\
142.9110	7.0211\\
155.8018	7.3114\\
169.2126	7.5993\\
183.1391	7.8847\\
199.0483	8.1958\\
215.5700	8.5038\\
232.6981	8.8088\\
250.4268	9.1108\\
270.4449	9.4367\\
291.1628	9.7591\\
308.8990	10.0219\\
321.5473	10.1948\\
334.4057	10.3577\\
349.3422	10.5317\\
364.5182	10.6927\\
379.9152	10.8407\\
395.5145	10.9757\\
411.2976	11.0977\\
427.2458	11.2068\\
443.3406	11.3028\\
459.5635	11.3858\\
475.8957	11.4558\\
492.3188	11.5128\\
508.8142	11.5569\\
525.3632	11.5879\\
541.9473	11.6059\\
558.5479	11.6110\\
575.1464	11.6030\\
591.7242	11.5821\\
608.2894	11.5480\\
624.8222	11.5009\\
641.2780	11.4407\\
657.6380	11.3674\\
673.8834	11.2810\\
689.9956	11.1815\\
705.9557	11.0690\\
721.7451	10.9434\\
737.3448	10.8047\\
752.7362	10.6529\\
767.9005	10.4880\\
782.8189	10.3101\\
795.6561	10.1437\\
808.2781	9.9672\\
820.6724	9.7808\\
832.8263	9.5843\\
844.7274	9.3778\\
856.3630	9.1613\\
867.7206	8.9348\\
878.7876	8.6983\\
888.0327	8.4875\\
897.0473	8.2695\\
910.1065	7.9483\\
918.5359	7.7593\\
926.7736	7.5911\\
933.5079	7.4668\\
940.1376	7.3570\\
946.6755	7.2617\\
953.1346	7.1808\\
959.5278	7.1144\\
965.8681	7.0624\\
972.1683	7.0249\\
978.4414	7.0018\\
984.7003	6.9932\\
990.9580	6.9991\\
997.2275	7.0194\\
1003.5215	7.0542\\
1009.8532	7.1034\\
1016.2353	7.1671\\
1022.6809	7.2452\\
1029.2029	7.3378\\
1035.8141	7.4449\\
1042.5276	7.5664\\
1049.3563	7.7024\\
1057.7210	7.8846\\
1066.2925	8.0877\\
1075.0932	8.3116\\
1084.1454	8.5563\\
1093.4715	8.8218\\
1104.7278	9.1579\\
1116.4229	9.5224\\
1128.5922	9.9151\\
1143.1260	10.3987\\
1160.3383	10.9870\\
1180.6216	11.6956\\
1206.6903	12.6103\\
1213.4961	12.8205\\
1220.4098	13.0136\\
1227.4223	13.1894\\
1234.5242	13.3480\\
1241.7063	13.4893\\
1248.9595	13.6134\\
1256.2745	13.7203\\
1263.6421	13.8099\\
1271.0530	13.8823\\
1278.4981	13.9375\\
1285.9680	13.9754\\
1293.4536	13.9960\\
1300.9457	13.9994\\
1308.4350	13.9856\\
1315.9122	13.9546\\
1323.3683	13.9063\\
1330.7939	13.8408\\
1338.1798	13.7580\\
1345.5168	13.6580\\
1352.7956	13.5407\\
1360.0071	13.4062\\
1367.1420	13.2545\\
1374.1911	13.0855\\
1381.1451	12.8993\\
1387.9948	12.6959\\
1394.7311	12.4752\\
1399.1543	12.3185\\
1403.5203	12.1541\\
1407.8262	11.9821\\
1412.0694	11.8024\\
1416.2472	11.6150\\
1420.3567	11.4200\\
1424.3952	11.2173\\
1428.3601	11.0070\\
1432.2486	10.7890\\
1436.0579	10.5633\\
1439.7853	10.3300\\
1443.4282	10.0890\\
1446.9836	9.8404\\
1450.4490	9.5841\\
1453.8216	9.3201\\
1457.0986	9.0485\\
1460.2774	8.7692\\
1463.3551	8.4822\\
1466.3291	8.1876\\
1469.1966	7.8853\\
1471.9549	7.5754\\
1474.6012	7.2578\\
1477.1328	6.9325\\
1479.5471	6.5996\\
1481.8411	6.2590\\
1484.0123	5.9107\\
1485.0510	5.7337\\
1486.0579	5.5548\\
1487.0327	5.3740\\
1487.9751	5.1912\\
1488.8847	5.0066\\
1489.7612	4.8200\\
1490.6042	4.6315\\
1491.4135	4.4411\\
1492.1886	4.2487\\
1492.9292	4.0545\\
1493.6350	3.8583\\
1494.3057	3.6603\\
1494.9408	3.4603\\
1495.5401	3.2584\\
1496.1032	3.0546\\
1496.6298	2.8488\\
1497.1195	2.6412\\
1497.5720	2.4316\\
1497.9869	2.2201\\
1498.3639	2.0068\\
1498.7027	1.7914\\
1499.0029	1.5742\\
1499.2642	1.3551\\
1499.4862	1.1340\\
1499.6686	0.9110\\
1499.8111	0.6862\\
1499.9133	0.4594\\
1499.9748	0.2306\\
1499.9954	0.0000\\
};
\addlegendentry{\scriptsize{No Traffic}}
\addplot [color=violet, dotted, line width=1.5pt]
  table[row sep=crcr]{%
0.0000	0.0000\\
0.0136	0.0722\\
0.0545	0.1442\\
0.1224	0.2159\\
0.2173	0.2874\\
0.3392	0.3586\\
0.4878	0.4296\\
0.6633	0.5003\\
0.9763	0.6060\\
1.3489	0.7111\\
1.7809	0.8156\\
2.2718	0.9196\\
2.8214	1.0230\\
3.6449	1.1600\\
4.5715	1.2960\\
5.6003	1.4310\\
6.7306	1.5651\\
7.9616	1.6982\\
9.6410	1.8631\\
11.4752	2.0265\\
13.4627	2.1883\\
15.6022	2.3486\\
17.8921	2.5074\\
20.8364	2.6958\\
23.9928	2.8820\\
27.3587	3.0660\\
30.9316	3.2478\\
35.3582	3.4571\\
40.0591	3.6633\\
45.0302	3.8665\\
50.2676	4.0667\\
55.7673	4.2638\\
62.3687	4.4854\\
69.3015	4.7030\\
76.5597	4.9167\\
84.1373	5.1264\\
93.0364	5.3576\\
102.3238	5.5838\\
111.9909	5.8050\\
122.0292	6.0211\\
133.6079	6.2554\\
145.6227	6.4835\\
158.0619	6.7055\\
170.9139	6.9212\\
185.5139	7.1514\\
200.5838	7.3741\\
216.1079	7.5894\\
232.0709	7.7971\\
249.9671	8.0152\\
268.3469	8.2244\\
287.1901	8.4246\\
306.4373	8.6156\\
327.6754	8.8117\\
349.3789	8.9975\\
371.5226	9.1730\\
394.0813	9.3381\\
418.8106	9.5044\\
443.9607	9.6587\\
469.5001	9.8011\\
495.3973	9.9315\\
521.6210	10.0499\\
550.0444	10.1634\\
578.7677	10.2632\\
607.7644	10.3493\\
637.0162	10.4217\\
666.4526	10.4804\\
696.0347	10.5252\\
725.7236	10.5563\\
755.4805	10.5735\\
785.2667	10.5770\\
815.0432	10.5667\\
844.7712	10.5426\\
874.4119	10.5047\\
907.8610	10.4465\\
929.4320	10.4177\\
950.9589	10.4038\\
972.4725	10.4048\\
994.0035	10.4207\\
1015.5828	10.4515\\
1037.2411	10.4972\\
1059.0094	10.5578\\
1080.9184	10.6333\\
1102.9989	10.7236\\
1125.2817	10.8289\\
1147.7976	10.9491\\
1170.5774	11.0842\\
1195.7654	11.2485\\
1208.4303	11.3302\\
1219.0161	11.3802\\
1229.6402	11.4125\\
1240.2863	11.4273\\
1250.9380	11.4245\\
1261.5789	11.4041\\
1272.1926	11.3662\\
1282.7627	11.3106\\
1293.2729	11.2375\\
1301.6268	11.1664\\
1309.9235	11.0840\\
1318.1546	10.9903\\
1326.3116	10.8854\\
1334.3862	10.7693\\
1342.3700	10.6419\\
1350.2546	10.5032\\
1358.0317	10.3533\\
1365.6927	10.1922\\
1373.2294	10.0198\\
1380.6334	9.8361\\
1387.8962	9.6412\\
1395.0094	9.4351\\
1401.9648	9.2177\\
1408.7538	8.9890\\
1415.3682	8.7491\\
1421.7994	8.4980\\
1426.4977	8.3023\\
1431.0847	8.1002\\
1435.5569	7.8918\\
1439.9108	7.6770\\
1444.1428	7.4560\\
1448.2494	7.2286\\
1452.2271	6.9949\\
1456.0723	6.7548\\
1459.7815	6.5085\\
1463.3511	6.2558\\
1466.7776	5.9968\\
1470.0575	5.7314\\
1473.1872	5.4597\\
1476.1632	5.1817\\
1478.0601	4.9928\\
1479.8860	4.8012\\
1481.6399	4.6067\\
1483.3208	4.4094\\
1484.9276	4.2093\\
1486.4593	4.0064\\
1487.9148	3.8006\\
1489.2931	3.5921\\
1490.5930	3.3807\\
1491.8137	3.1666\\
1492.9539	2.9496\\
1494.0128	2.7298\\
1494.9892	2.5072\\
1495.8820	2.2818\\
1496.6903	2.0535\\
1497.4129	1.8225\\
1497.7418	1.7059\\
1498.0489	1.5886\\
1498.3341	1.4706\\
1498.5972	1.3520\\
1498.8381	1.2326\\
1499.0567	1.1125\\
1499.2528	0.9917\\
1499.4264	0.8702\\
1499.5772	0.7480\\
1499.7052	0.6251\\
1499.8102	0.5015\\
1499.8921	0.3772\\
1499.9507	0.2521\\
1499.9860	0.1264\\
1499.9978	0.0000\\
};
\addlegendentry{\scriptsize{Low Speed+No Traffic}}
\end{axis}%
\end{tikzpicture}%

%% file: Images/scenario.tex
\begin{tikzpicture}%
\node[anchor=south west,inner sep=0] (Scenario) at (0,0) {\includegraphics[width=1.00\columnwidth]{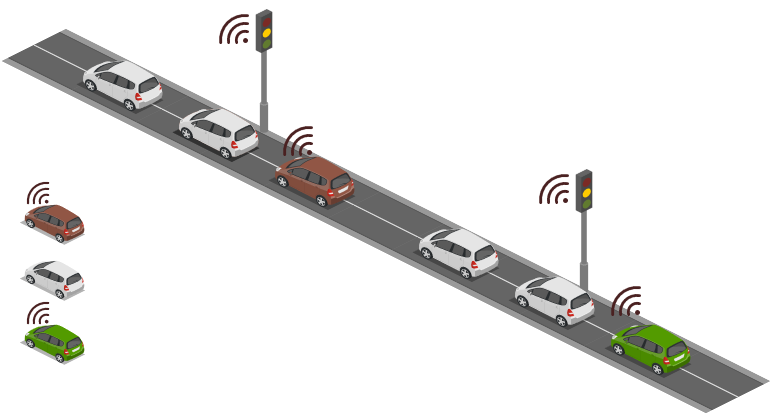}};
\node[anchor = west] at (1.10, 2.20) {\small{Virtual (C)AV}};
\node[anchor = west] at (1.10, 1.55) {\small{Virtual Human}};
\node[anchor = west] at (1.11, 0.80) {\small{Physical Ego Vehicle}};
\node[anchor = north west, font = \scriptsize] at (6.65, 4.85) {V2I Range};
\node [anchor = north west, font=\scriptsize] at (6.85, 4.45) {$-$ \unit[0]{km}};
\node [anchor = north west, font=\scriptsize] at (6.85, 4.05)  {$-$ \unit[0.15]{km}};
\node [anchor = north west, font=\scriptsize] at (6.85, 3.65)  {$-$ \unit[0.30]{km}};
\node [anchor = north west, font=\scriptsize] at (6.85, 3.25)  {$-$ \unit[0.60]{km}};
\node [anchor = north west, font=\scriptsize] at (6.85, 2.85)  {$-$ \unit[1.20]{km}};
\node[anchor = north west, font = \scriptsize] at (3.35, 4.85) {Downstream Traffic};
\node [anchor = north west, font=\scriptsize] at (3.55, 4.45) {$-$ No CAVs};
\node [anchor = north west, font=\scriptsize] at (3.55, 4.05)  {$-$ Partial (C)AVs};
\end{tikzpicture}%

%% file: Images/peachtree.tex
%
%
\definecolor{mycolor1}{rgb}{1.00000,1.00000,0.00000}%
\begin{tikzpicture}

\begin{axis}[%
width=3.7cm,
height=0.507\columnwidth,
at={(-3.7cm,0cm)},
scale only axis,
unbounded coords=jump,
xmin=0,
xmax=250,
xtick={  0,  50, 100, 150, 200, 250, 300},
xlabel style={font=\color{white!15!black},font=\small},
xlabel={Time [s]},
ymin=-0.1,
ymax=1570,
ytick={0,  250,  500,  750, 1000, 1250, 1500, 1750, 2000, 2250, 2500, 2750, 3000},
yticklabels={0, 0.25, 0.50, 0.75, 1.00, 1.25, 1.50, 1.75, 2.00, 2.25, 2.50, 2.75, 3.00},
ylabel style={font=\color{white!15!black},font=\small},
ylabel={Position [km]},
axis background/.style={fill=white},
xmajorgrids,
ticklabel style={font=\scriptsize}
]
\addplot [color=green, line width=2.0pt, forget plot]
  table[row sep=crcr]{%
35	170\\
115	170\\
nan	nan\\
160	170\\
240	170\\
};
\addplot [color=mycolor1, line width=2.0pt, forget plot]
  table[row sep=crcr]{%
116	170\\
120	170\\
nan	nan\\
241	170\\
245	170\\
};
\addplot [color=red, line width=2.0pt, forget plot]
  table[row sep=crcr]{%
0	170\\
34	170\\
nan	nan\\
121	170\\
159	170\\
nan	nan\\
246	170\\
250	170\\
};
\addplot [color=green, line width=2.0pt, forget plot]
  table[row sep=crcr]{%
3	460\\
28	460\\
nan	nan\\
98	460\\
123	460\\
nan	nan\\
193	460\\
218	460\\
};
\addplot [color=mycolor1, line width=2.0pt, forget plot]
  table[row sep=crcr]{%
29	460\\
31	460\\
nan	nan\\
124	460\\
126	460\\
nan	nan\\
219	460\\
222	460\\
};
\addplot [color=red, line width=2.0pt, forget plot]
  table[row sep=crcr]{%
0	460\\
2	460\\
nan	nan\\
32	460\\
97	460\\
nan	nan\\
127	460\\
192	460\\
nan	nan\\
223	460\\
250	460\\
};
\addplot [color=green, line width=2.0pt, forget plot]
  table[row sep=crcr]{%
1	620\\
36	620\\
nan	nan\\
96	620\\
131	620\\
nan	nan\\
191	620\\
226	620\\
};
\addplot [color=mycolor1, line width=2.0pt, forget plot]
  table[row sep=crcr]{%
37	620\\
40	620\\
nan	nan\\
132	620\\
135	620\\
nan	nan\\
227	620\\
230	620\\
};
\addplot [color=red, line width=2.0pt, forget plot]
  table[row sep=crcr]{%
0	620\\
nan	nan\\
41	620\\
95	620\\
nan	nan\\
136	620\\
190	620\\
nan	nan\\
231	620\\
250	620\\
};
\addplot [color=green, line width=2.0pt, forget plot]
  table[row sep=crcr]{%
0	780\\
36	780\\
nan	nan\\
75	780\\
131	780\\
nan	nan\\
170	780\\
226	780\\
};
\addplot [color=mycolor1, line width=2.0pt, forget plot]
  table[row sep=crcr]{%
37	780\\
39	780\\
nan	nan\\
132	780\\
134	780\\
nan	nan\\
227	780\\
229	780\\
};
\addplot [color=red, line width=2.0pt, forget plot]
  table[row sep=crcr]{%
40	780\\
74	780\\
nan	nan\\
135	780\\
169	780\\
nan	nan\\
230	780\\
250	780\\
};
\addplot [color=green, line width=2.0pt, forget plot]
  table[row sep=crcr]{%
0	1020\\
17	1020\\
nan	nan\\
83	1020\\
110	1020\\
nan	nan\\
176	1020\\
203	1020\\
};
\addplot [color=mycolor1, line width=2.0pt, forget plot]
  table[row sep=crcr]{%
18	1020\\
20	1020\\
nan	nan\\
111	1020\\
113	1020\\
nan	nan\\
204	1020\\
206	1020\\
};
\addplot [color=red, line width=2.0pt, forget plot]
  table[row sep=crcr]{%
21	1020\\
82	1020\\
nan	nan\\
114	1020\\
175	1020\\
nan	nan\\
207	1020\\
250	1020\\
};
\addplot [color=red, line width=2.0pt, forget plot]
  table[row sep=crcr]{%
0	1420\\
250	1420\\
};
\end{axis}
\node[anchor=south west,inner sep=0] (peachtree) at (1,-0.5cm) {\includegraphics[width=0.30\columnwidth]{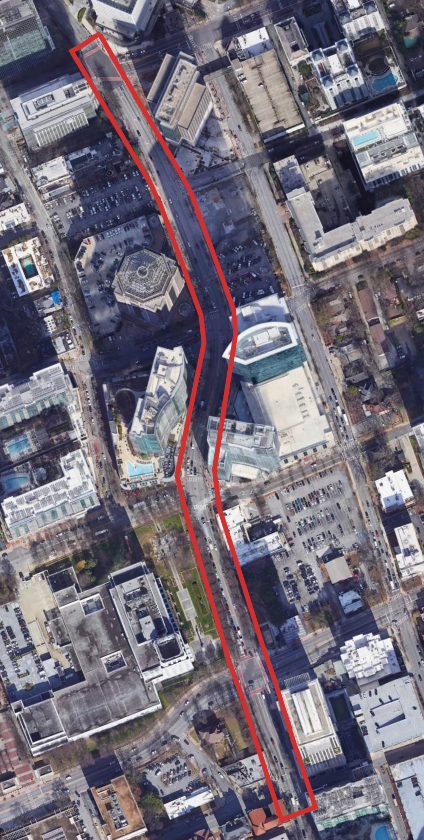}};
\draw (peachtree.north east) rectangle (peachtree.south west);
\draw [draw=black, fill=green, opacity=0.1]
       (-3.80, 1) -- (-3.80, 3.2) -- (0.1, 3.2) -- (0.1, 1) -- cycle;
\draw [thick, draw=green, opacity=0.5] (0.1, 3.2) -- (peachtree.north west);
\draw [thick, draw=green, opacity=0.5] (0.1, 1) -- (peachtree.south west);
\end{tikzpicture}%

%% file: Images/hvbaseline.tex
\begin{tikzpicture}[font={\scriptsize}]
\node[anchor=south west, inner sep=0] (HVbaseline) at (0,0) {\includegraphics[height=0.35\linewidth,keepaspectratio]{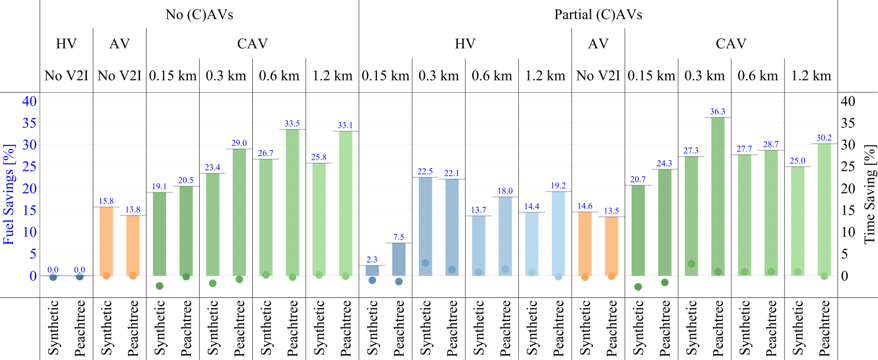}};
\node[anchor = east, align=right] at (0.50, 6.10) {Downstream Traffic};
\node[anchor = east, align=right] at (0.50, 5.57) {Ego Controller};
\node[anchor = east, align=right] at (0.50, 5.05) {V2I Range};
\node[anchor = east, align=right] at (0.50, 0.60) {Route};
\end{tikzpicture}

%% file: Images/avbaseline.tex
\begin{tikzpicture}[font={\scriptsize}]
\node[anchor=south west, inner sep=0] (AVbaseline) at (0,0) {\includegraphics[height=0.32\linewidth,keepaspectratio]{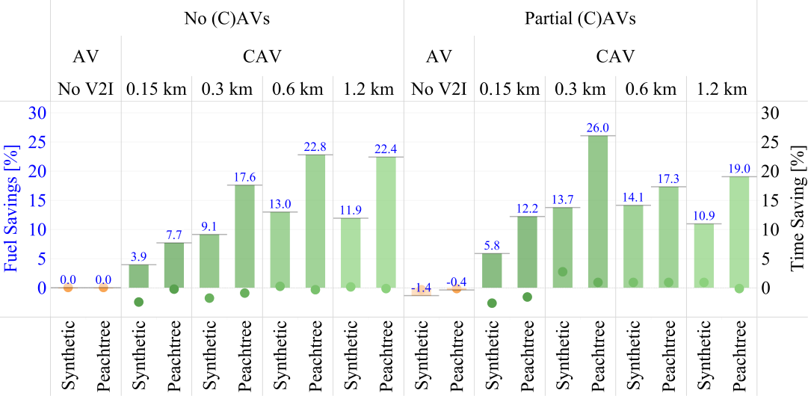}};
\node[anchor = east, align=right] at (0.50, 5.50) {Downstream Traffic};
\node[anchor = east, align=right] at (0.50, 5.00) {Ego Controller};
\node[anchor = east, align=right] at (0.50, 4.50) {V2I Range};
\node[anchor = east, align=right] at (0.50, 0.60) {Route};
\end{tikzpicture}

%% file: Images/autonomie.tex
\begin{tikzpicture}[scale=1, transform shape]
    \pgfmathsetmacro{\gconv}{1.31992} 
    \begin{axis}[
    view={110}{20},
    width=0.957\columnwidth,
    height=0.757\columnwidth,
    grid=major,
    xmin=-1,xmax=5,
    ymin=0,ymax=15,
    zmin=0,zmax=150,
    xtick={4,3,2,1,0},
    xticklabels={0, 0.15, 0.30, 0.60, 1.20},
    ytick={1,2,3,4,5,6,7,8,9,10,11,12,13,14},
    yticklabels={1, , 3, , 5, , 7, , 9, , 11, , 13, \text{mean}},
    ztick={20, 45, 70, 95, 120, 145},
    zticklabels={0, 25, 50, 75, 100, 125},
    xlabel={V2I Range [km]},
    ylabel={Vehicle ID},
    zlabel={Fuel Improvement [\%]},
    xlabel style={sloped,font=\small},
    ylabel style={sloped,font=\small},
    zlabel style={sloped,font=\small},
    ticklabel style={font=\scriptsize},
    xticklabel style={rotate=-20},
    yticklabel style={rotate=38},
    xlabel shift = -5 pt,
    ]
    \path let \p1=($(axis cs:0,0,1)-(axis cs:0,0,0)$) in 
    \pgfextra{\pgfmathsetmacro{\conv}{2*\y1}
    \ifx\gconv\conv
    \typeout{z-scale\space good!}
    \else
    \typeout{Kindly\space consider\space setting\space the\space 
            prefactor\space of\space z\space to\space \conv}
    \fi     
            };  
    
    \pgfplotsset{3d bars/.style={only marks,scatter,mark=half cube*,mark size=0.3cm, 
    3d cube color=#1,point meta=0,
    ,visualization depends on={\gconv*z \as \myz},
    scatter/@pre marker code/.append style={/pgfplots/cube/size z=\myz},}}
    
    
    
    
    
    
    \addplot3[3d bars=green] 
    coordinates {(0,1,156.610) (0,2,83.068) (0,3,113.058) (0,4,68.186) (0,5,59.629) (0,6,39.119) (0,7,42.676) (0,8,31.852) (0,9,36.460) (0,10,40.544) (0,11,60.617) (0,12,55.926) (0,13,47.463) (0,14,64.247) };    
    \addplot3[3d bars=blue] 
    coordinates {(1,1,157.261) (1,2,83.776) (1,3,105.777) (1,4,67.728) (1,5,54.926) (1,6,43.633) (1,7,47.257) (1,8,34.943) (1,9,30.125) (1,10,37.546) (1,11,67.909) (1,12,56.577) (1,13,49.700) (1,14,64.397) };    
    \addplot3[3d bars=orange] 
    coordinates {(2,1,138.278) (2,2,54.785) (2,3,67.530) (2,4,52.192) (2,5,48.301) (2,6,50.521) (2,7,50.452) (2,8,39.145) (2,9,42.511) (2,10,46.901) (2,11,83.423) (2,12,76.163) (2,13,62.082) (2,14,62.483) };    
    \addplot3[3d bars=violet] 
    coordinates {(3,1,74.213) (3,2,52.192) (3,3,60.794) (3,4,36.906) (3,5,25.808) (3,6,18.269) (3,7,9.596) (3,8,2.995) (3,9,17.095) (3,10,8.432) (3,11,58.811) (3,12,50.922) (3,13,45.759) (3,14,35.232) };
    \addplot3[3d bars=red] 
    coordinates {(4,1,16.057) (4,2,6.706) (4,3,23.203) (4,4,3.415) (4,5,15.465) (4,6,16.076) (4,7,20.689) (4,8,12.631) (4,9,18.576) (4,10,17.462) (4,11,40.212) (4,12,37.785) (4,13,24.937) (4,14,19.149) };
    \end{axis}%
\end{tikzpicture}%

%% file: Images/peachtreeautonomie.tex
\begin{tikzpicture}[scale=1, transform shape]
    \pgfmathsetmacro{\gconv}{1.31992} 
    \begin{axis}[
    view={110}{20},
    width=0.957\columnwidth,
    height=0.757\columnwidth,
    grid=major,
    xmin=-1,xmax=5,
    ymin=0,ymax=15,
    zmin=-0,zmax=150,
    xtick={4,3,2,1,0},
    xticklabels={0, 0.15, 0.30, 0.60, 1.20},
    ytick={1,2,3,4,5,6,7,8,9,10,11,12,13,14},
    yticklabels={1, , 3, , 5, , 7, , 9, , 11, , 13, \text{mean}},
    ztick={20, 45, 70, 95, 120, 145},
    zticklabels={0, 25, 50, 75, 100, 125},
    xlabel={V2I Range [km]},
    ylabel={Vehicle ID},
    zlabel={Fuel Improvement [\%]},
    xlabel style={sloped,font=\small},
    ylabel style={sloped,font=\small},
    zlabel style={sloped,font=\small},
    ticklabel style={font=\scriptsize},
    xticklabel style={rotate=-20},
    yticklabel style={rotate=38},
    xlabel shift = -5 pt,
    ]

    \path let \p1=($(axis cs:0,0,1)-(axis cs:0,0,0)$) in 
    \pgfextra{\pgfmathsetmacro{\conv}{2*\y1}
    \ifx\gconv\conv
    \typeout{z-scale\space good!}
    \else
    \typeout{Kindly\space consider\space setting\space the\space 
            prefactor\space of\space z\space to\space \conv}
    \fi     
            };  
    
    \pgfplotsset{3d bars/.style={only marks,scatter,mark=half cube*,mark size=0.3cm, 
    3d cube color=#1,point meta=0,
    ,visualization depends on={\gconv*z \as \myz},
    scatter/@pre marker code/.append style={/pgfplots/cube/size z=\myz},}}
    
    \addplot3[3d bars=green] 
coordinates {(0,1,91.253) (0,2,80.456) (0,3,88.106) (0,4,66.359) (0,5,70.498) (0,6,51.804) (0,7,50.547) (0,8,48.230) (0,9,79.540) (0,10,62.561) (0,11,75.904) (0,12,50.689) (0,13,49.899) (0,14,66.604) };    
    \addplot3[3d bars=blue] 
coordinates {(1,1,90.527) (1,2,82.555) (1,3,80.466) (1,4,77.382) (1,5,91.467) (1,6,64.630) (1,7,57.824) (1,8,56.250) (1,9,81.928) (1,10,64.072) (1,11,92.637) (1,12,65.431) (1,13,59.070) (1,14,74.172) };    
    \addplot3[3d bars=orange] 
coordinates {(2,1,93.843) (2,2,74.989) (2,3,75.208) (2,4,65.398) (2,5,68.577) (2,6,58.706) (2,7,55.404) (2,8,51.713) (2,9,90.120) (2,10,54.375) (2,11,95.679) (2,12,66.917) (2,13,64.589) (2,14,70.424) };    
    \addplot3[3d bars=violet] 
coordinates {(3,1,46.658) (3,2,48.615) (3,3,51.450) (3,4,50.413) (3,5,45.792) (3,6,25.856) (3,7,24.450) (3,8,23.074) (3,9,23.741) (3,10,24.198) (3,11,61.580) (3,12,51.416) (3,13,59.852) (3,14,41.315) };    
    \addplot3[3d bars=red] 
coordinates {(4,1,13.440) (4,2,16.181) (4,3,22.258) (4,4,23.567) (4,5,18.950) (4,6,19.654) (4,7,19.738) (4,8,21.490) (4,9,20.409) (4,10,22.909) (4,11,28.866) (4,12,19.226) (4,13,19.804) (4,14,20.116) };    
    \end{axis}%
\end{tikzpicture}%

%% file: root.bbl
\begin{thebibliography}{10}
\providecommand{\url}[1]{#1}
\csname url@samestyle\endcsname
\providecommand{\newblock}{\relax}
\providecommand{\bibinfo}[2]{#2}
\providecommand{\BIBentrySTDinterwordspacing}{\spaceskip=0pt\relax}
\providecommand{\BIBentryALTinterwordstretchfactor}{4}
\providecommand{\BIBentryALTinterwordspacing}{\spaceskip=\fontdimen2\font plus
\BIBentryALTinterwordstretchfactor\fontdimen3\font minus
  \fontdimen4\font\relax}
\providecommand{\BIBforeignlanguage}[2]{{%
\expandafter\ifx\csname l@#1\endcsname\relax
\typeout{** WARNING: IEEEtran.bst: No hyphenation pattern has been}%
\typeout{** loaded for the language `#1'. Using the pattern for}%
\typeout{** the default language instead.}%
\else
\language=\csname l@#1\endcsname
\fi
#2}}
\providecommand{\BIBdecl}{\relax}
\BIBdecl

\bibitem{Vahidi2018}
\BIBentryALTinterwordspacing
A.~Vahidi and A.~Sciarretta, ``{Energy saving potentials of connected and
  automated vehicles},'' \emph{Transportation Research Part C: Emerging
  Technologies}, vol.~95, no. April 2018, pp. 822--843, 2018. [Online].
  Available: \url{https://doi.org/10.1016/j.trc.2018.09.001}
\BIBentrySTDinterwordspacing

\bibitem{Guanetti2018}
\BIBentryALTinterwordspacing
J.~Guanetti, Y.~Kim, and F.~Borrelli, ``{Control of connected and automated
  vehicles: State of the art and future challenges},'' \emph{Annual Reviews in
  Control}, vol.~45, no. May, pp. 18--40, 2018. [Online]. Available:
  \url{https://doi.org/10.1016/j.arcontrol.2018.04.011}
\BIBentrySTDinterwordspacing

\bibitem{Wang2020}
Z.~Wang, Y.~Bian, S.~E. Shladover, G.~Wu, S.~E. Li, and M.~J. Barth, ``{A
  Survey on Cooperative Longitudinal Motion Control of Multiple Connected and
  Automated Vehicles},'' \emph{IEEE Intelligent Transportation Systems
  Magazine}, vol.~12, no.~1, pp. 4--24, 2020.

\bibitem{Sarkar2016}
\BIBentryALTinterwordspacing
R.~Sarkar and J.~Ward, \emph{DOE SMART Mobility: Systems and Modeling for
  Accelerated Research in Transportation}.\hskip 1em plus 0.5em minus
  0.4em\relax Cham: Springer International Publishing, 2016, pp. 39--52.
  [Online]. Available: \url{https://doi.org/10.1007/978-3-319-40503-2_4}
\BIBentrySTDinterwordspacing

\bibitem{Shladover2015}
S.~E. Shladover, C.~Nowakowski, X.-Y. Lu, and R.~Ferlis, ``Cooperative adaptive
  cruise control: Definitions and operating concepts,'' \emph{Transportation
  Research Record}, vol. 2489, 2015.

\bibitem{Asadi2011}
B.~Asadi and A.~Vahidi, ``{Predictive cruise control: Utilizing upcoming
  traffic signal information for improving fuel economy and reducing trip
  time},'' \emph{IEEE Transactions on Control Systems Technology}, vol.~19,
  no.~3, pp. 707--714, 2011.

\bibitem{Talebpour2016}
\BIBentryALTinterwordspacing
A.~Talebpour and H.~S. Mahmassani, ``{Influence of connected and autonomous
  vehicles on traffic flow stability and throughput},'' \emph{Transportation
  Research Part C: Emerging Technologies}, vol.~71, no. July, pp. 143--163,
  2016. [Online]. Available: \url{http://dx.doi.org/10.1016/j.trc.2016.07.007}
\BIBentrySTDinterwordspacing

\bibitem{Rios-Torres2018}
J.~Rios-Torres and A.~A. Malikopoulos, ``{Impact of Partial Penetrations of
  Connected and Automated Vehicles on Fuel Consumption and Traffic Flow},''
  \emph{IEEE Transactions on Intelligent Vehicles}, vol.~3, no.~4, pp.
  453--462, 2018.

\bibitem{ARD2020b}
\BIBentryALTinterwordspacing
T.~Ard, R.~A. Dollar, A.~Vahidi, Y.~Zhang, and D.~Karbowski, ``Microsimulation
  of energy and flow effects from optimal automated driving in mixed traffic,''
  \emph{Transportation Research Part C: Emerging Technologies}, vol. 120, p.
  102806, 2020. [Online]. Available:
  \url{https://www.sciencedirect.com/science/article/pii/S0968090X20307130}
\BIBentrySTDinterwordspacing

\bibitem{Liu2020}
H.~Liu, S.~E. Shladover, X.~Y. Lu, and X.~Kan, ``{Freeway vehicle fuel
  efficiency improvement via cooperative adaptive cruise control},''
  \emph{Journal of Intelligent Transportation Systems: Technology, Planning,
  and Operations}, vol.~0, no.~0, pp. 1--13, 2020.

\bibitem{Jayawardana2022}
V.~Jayawardana and C.~Wu, ``Learning eco-driving strategies at signalized
  intersections,'' in \emph{European Control Conference (ECC)}, 2022.

\bibitem{WAN2016}
\BIBentryALTinterwordspacing
N.~Wan, A.~Vahidi, and A.~Luckow, ``Optimal speed advisory for connected
  vehicles in arterial roads and the impact on mixed traffic,''
  \emph{Transportation Research Part C: Emerging Technologies}, vol.~69, pp.
  548--563, 2016. [Online]. Available:
  \url{https://www.sciencedirect.com/science/article/pii/S0968090X16000292}
\BIBentrySTDinterwordspacing

\bibitem{Yang2017}
H.~Yang, H.~Rakha, and M.~V. Ala, ``Eco-cooperative adaptive cruise control at
  signalized intersections considering queue effects,'' \emph{IEEE Transactions
  on Intelligent Transportation Systems}, vol.~18, no.~6, pp. 1575--1585, 2017.

\bibitem{Ge2018}
\BIBentryALTinterwordspacing
J.~I. Ge, S.~S. Avedisov, C.~R. He, W.~B. Qin, M.~Sadeghpour, and G.~Orosz,
  ``{Experimental validation of connected automated vehicle design among
  human-driven vehicles},'' \emph{Transportation Research Part C: Emerging
  Technologies}, vol.~91, no. April, pp. 335--352, 2018. [Online]. Available:
  \url{https://doi.org/10.1016/j.trc.2018.04.005}
\BIBentrySTDinterwordspacing

\bibitem{Avedisov2022}
S.~S. Avedisov, G.~Bansal, and G.~Orosz, ``Impacts of connected automated
  vehicles on freeway traffic patterns at different penetration levels,''
  \emph{IEEE Transactions on Intelligent Transportation Systems}, vol.~23,
  no.~5, pp. 4305--4318, 2022.

\bibitem{Ibrahim2019}
A.~Ibrahim, M.~Cicic, D.~Goswami, T.~Basten, and K.~H. Johansson, ``{Control of
  Platooned Vehicles in Presence of Traffic Shock Waves},'' \emph{2019 IEEE
  Intelligent Transportation Systems Conference, ITSC 2019}, pp. 1727--1734,
  2019.

\bibitem{Mahler2017}
G.~Mahler, A.~Winckler, S.~A. Fayazi, M.~Filusch, and A.~Vahidi, ``Cellular
  communication of traffic signal state to connected vehicles for arterial
  eco-driving,'' in \emph{2017 IEEE 20th International Conference on
  Intelligent Transportation Systems (ITSC)}, 2017, pp. 1--6.

\bibitem{Chen2022}
\BIBentryALTinterwordspacing
H.~Chen and H.~A. Rakha, ``Developing and field testing a green light optimal
  speed advisory system for buses,'' \emph{Energies}, vol.~15, no.~4, 2022.
  [Online]. Available: \url{https://www.mdpi.com/1996-1073/15/4/1491}
\BIBentrySTDinterwordspacing

\bibitem{Kamal2011}
\BIBentryALTinterwordspacing
M.~Kamal, M.~Mukai, J.~Murata, and T.~Kawabe, ``Ecological driving based on
  preceding vehicle prediction using {MPC},'' \emph{IFAC Proceedings Volumes},
  vol.~44, no.~1, pp. 3843--3848, 2011, 18th IFAC World Congress. [Online].
  Available:
  \url{https://www.sciencedirect.com/science/article/pii/S1474667016442102}
\BIBentrySTDinterwordspacing

\bibitem{kamal2013}
M.~A.~S. Kamal, M.~Mukai, J.~Murata, and T.~Kawabe, ``Model predictive control
  of vehicles on urban roads for improved fuel economy,'' \emph{IEEE
  Transactions on Control Systems Technology}, vol.~21, no.~3, pp. 831--841,
  2013.

\bibitem{Wang2020b}
Z.~Wang, G.~Wu, and M.~J. Barth, ``Cooperative eco-driving at signalized
  intersections in a partially connected and automated vehicle environment,''
  \emph{IEEE Transactions on Intelligent Transportation Systems}, vol.~21,
  no.~5, pp. 2029--2038, 2020.

\bibitem{HomChaudhuri2017}
B.~HomChaudhuri, A.~Vahidi, and P.~Pisu, ``{Fast Model Predictive Control-Based
  Fuel Efficient Control Strategy for a Group of Connected Vehicles in Urban
  Road Conditions},'' \emph{IEEE Transactions on Control Systems Technology},
  vol.~25, no.~2, pp. 760--767, 2017.

\bibitem{Zhang2016}
Y.~J. Zhang, A.~A. Malikopoulos, and C.~G. Cassandras, ``{Optimal control and
  coordination of connected and automated vehicles at urban traffic
  intersections},'' \emph{Proceedings of the American Control Conference}, vol.
  2016-July, pp. 6227--6232, 2016.

\bibitem{Malikopoulos2018}
\BIBentryALTinterwordspacing
A.~A. Malikopoulos, C.~G. Cassandras, and Y.~J. Zhang, ``{A decentralized
  energy-optimal control framework for connected automated vehicles at
  signal-free intersections},'' \emph{Automatica}, vol.~93, pp. 244--256, 2018.
  [Online]. Available: \url{https://doi.org/10.1016/j.automatica.2018.03.056}
\BIBentrySTDinterwordspacing

\bibitem{Mahler2014}
G.~Mahler and A.~Vahidi, ``An optimal velocity-planning scheme for vehicle
  energy efficiency through probabilistic prediction of traffic-signal
  timing,'' \emph{IEEE Transactions on Intelligent Transportation Systems},
  vol.~15, no.~6, pp. 2516--2523, 2014.

\bibitem{Sun2020}
C.~Sun, J.~Guanetti, F.~Borrelli, and S.~J. Moura, ``{Optimal Eco-Driving
  Control of Connected and Autonomous Vehicles Through Signalized
  Intersections},'' \emph{IEEE Internet of Things Journal}, vol.~7, no.~5, pp.
  3759--3773, 2020.

\bibitem{Dresner2008}
K.~Dresner and P.~Stone, ``{A Multiagent Approach to Autonomous Intersection
  Management},'' \emph{Journal of Artificial Intelligence Research}, 2008.

\bibitem{Quinlan2010}
M.~Quinlan, T.~C. Au, J.~Zhu, N.~Stiurca, and P.~Stone, ``{Bringing simulation
  to life: A mixed reality autonomous intersection},'' \emph{IEEE/RSJ 2010
  International Conference on Intelligent Robots and Systems, IROS 2010 -
  Conference Proceedings}, no. October, pp. 6083--6088, 2010.

\bibitem{Fayazi2017}
S.~A. Fayazi and A.~Vahidi, ``Vehicle-in-the-loop (vil) verification of a smart
  city intersection control scheme for autonomous vehicles,'' in \emph{2017
  IEEE Conference on Control Technology and Applications (CCTA)}, 2017, pp.
  1575--1580.

\bibitem{MalikopoulosVIL}
\BIBentryALTinterwordspacing
A.~M.~I. Mahbub, V.~Karri, D.~Parikh, S.~Jade, and A.~A. Malikopoulos, ``A
  decentralized time- and energy-optimal control framework for connected
  automated vehicles: From simulation to field test,'' in \emph{SAE Technical
  Paper}.\hskip 1em plus 0.5em minus 0.4em\relax SAE International, 04 2020.
  [Online]. Available: \url{https://doi.org/10.4271/2020-01-0579}
\BIBentrySTDinterwordspacing

\bibitem{FengAR}
Y.~{Feng}, C.~{Yu}, S.~{Xu}, H.~X. {Liu}, and H.~{Peng}, ``An augmented reality
  environment for connected and automated vehicle testing and evaluation,'' in
  \emph{2018 IEEE Intelligent Vehicles Symposium (IV)}, 2018, pp. 1549--1554.

\bibitem{Ard2021}
\BIBentryALTinterwordspacing
T.~Ard, L.~Guo, R.~A. Dollar, A.~Fayazi, N.~Goulet, Y.~Jia, B.~Ayalew, and
  A.~Vahidi, ``Energy and flow effects of optimal automated driving in mixed
  traffic: Vehicle-in-the-loop experimental results,'' \emph{Transportation
  Research Part C: Emerging Technologies}, vol. 130, p. 103168, 2021. [Online].
  Available:
  \url{https://www.sciencedirect.com/science/article/pii/S0968090X21001868}
\BIBentrySTDinterwordspacing

\bibitem{Xia2012}
H.~Xia, K.~Boriboonsomsin, F.~Schweizer, A.~Winckler, K.~Zhou, W.-B. Zhang, and
  M.~Barth, ``Field operational testing of eco-approach technology at a
  fixed-time signalized intersection,'' in \emph{2012 15th International IEEE
  Conference on Intelligent Transportation Systems}, 2012, pp. 188--193.

\bibitem{Almannaa2019}
\BIBentryALTinterwordspacing
M.~H. Almannaa, H.~Chen, H.~A. Rakha, A.~Loulizi, and I.~El-Shawarby, ``Field
  implementation and testing of an automated eco-cooperative adaptive cruise
  control system in the vicinity of signalized intersections,''
  \emph{Transportation Research Part D: Transport and Environment}, vol.~67,
  pp. 244--262, 2019. [Online]. Available:
  \url{https://www.sciencedirect.com/science/article/pii/S1361920918305583}
\BIBentrySTDinterwordspacing

\bibitem{Bae2022}
\BIBentryALTinterwordspacing
S.~Bae, Y.~Kim, Y.~Choi, J.~Guanetti, P.~Gill, F.~Borrelli, and S.~J. Moura,
  ``Ecological adaptive cruise control of plug-in hybrid electric vehicle with
  connected infrastructure and on-road experiments,'' \emph{Journal of Dynamic
  Systems, Measurement, and Control}, vol. 144, no.~1, 01 2022, 011109.
  [Online]. Available: \url{https://doi.org/10.1115/1.4053187}
\BIBentrySTDinterwordspacing

\bibitem{Han2021}
J.~Han, D.~Shen, D.~Karbowski, and A.~Rousseau, ``{Leveraging multiple
  connected traffic light signals in an energy-efficient speed planner},''
  \emph{IEEE Control Systems Letters}, vol.~5, no.~6, pp. 2078--2083, 2021.

\bibitem{Bryson}
A.~E. Bryson and Y.~Ho, \emph{Applied optimal control: optimization,
  estimation, and control}.\hskip 1em plus 0.5em minus 0.4em\relax CRC Press,
  1975.

\bibitem{ptvvissim10}
{PTV Group}, \emph{{PTV VISSIM 10 User Manual}}.\hskip 1em plus 0.5em minus
  0.4em\relax {PTV AG}, 2018.

\bibitem{autonomie}
S.~Halbach, P.~Sharer, S.~Pagerit, C.~Folkerts, and A.~Rousseau, ``Model
  architecture, methods, and interfaces for efficient math-based design and
  simulation of automotive control systems,'' 04 2010.

\end{thebibliography}
